\documentclass[journal,comsoc]{IEEEtran}

\usepackage[T1]{fontenc}
  
\ifCLASSINFOpdf
  \usepackage[pdftex]{graphicx}
\else

\fi                   

\usepackage{amsmath}    

\usepackage[cmintegrals]{newtxmath}

\usepackage{booktabs}

\makeatletter
\def\ps@IEEEtitlepagestyle{%
  \def\@oddfoot{\mycopyrightnotice}%
  \def\@evenfoot{}%
}
\def\mycopyrightnotice{%
{\footnotesize
  \begin{minipage}{\textwidth}
  Copyright~\copyright~2021 IEEE. Personal use of this material is permitted. However, permission to use this material for any other purposes must be obtained from the IEEE by sending a request to pubs-permissions@ieee.org.
  \end{minipage}
  }
}
\makeatother

\begin{document}

\title{Leader Confirmation Replication for Millisecond Consensus in Private Chains}
\author{Haiwen~Du,~
        Dongjie~Zhu,~
        Yundong~Sun~
        and~Zhaoshuo~Tian,~\IEEEmembership{Member,~IEEE}
       
\IEEEcompsocitemizethanks{\IEEEcompsocthanksitem D. Zhu is with the School
of Computer Science and Technology, Harbin Institute of Technology, Weihai,
Shandong, China, 264209. (E-mail: zhudongjie@hit.edu.cn)
\IEEEcompsocthanksitem H. Du, Y. Sun and Z. Tian are with the School
of Astronautics, Harbin Institute of Technology, Harbin, China, 150001.\\
}
\thanks{Manuscript accepted Sep. 15, 2021}
\thanks{(Corresponding author: Dongjie Zhu.)}
}                 
 
\markboth{IEEE Internet of Things Journal}
{Shell \MakeLowercase{\textit{et al.}}: Bare Demo of IEEEtran.cls for Computer Society Journals}

\maketitle

\begin{abstract}
    The private chain-based Internet of Things (IoT) system ensures the security of cross-organizational data sharing. As a widely used consensus model in private chains, the leader-based state-machine replication (SMR) model meets the performance bottleneck in IoT blockchain applications, where nontransactional sensor data are generated on a scale. We analyzed IoT private chain systems and found that the leader maintains too many connections due to high latency and client request frequency, which results in lower consensus performance and efficiency.
    
    To meet this challenge, we propose a novel solution for maintaining low request latency and high transactions per second (TPS): replicate nontransactional data by followers and confirm by the leader to achieve nonconfliction SMR, rather than all by the leader. Our solution, named Leader Confirmation Replication (LCR), uses the newly proposed future log and confirmation signal to achieve nontransactional data replication on the followers, thereby reducing the leader’s network traffic and the request latency of transactional data. In addition, the generation replication strategy is designed to ensure the reliability and consistency of LCR when meeting membership changes. We evaluated LCR with various cluster sizes and network latencies. Experimental results show that in ms-network latency (2-30) environments, the TPS of LCR is 1.4X-1.9X higher than Raft, the transactional data response time is reduced by 40\%-60\%, and the network traffic is reduced by 20\%-30\% with acceptable network traffic and CPU cost on the followers. In addition, LCR shows high portability and availability since it does not change the number of leaders or the election process.
\end{abstract}

\begin{IEEEkeywords}
    confirmation replication,
    state-machine replication,
    consensus model,
    Raft,
    private chain
\end{IEEEkeywords}
        
\IEEEpeerreviewmaketitle

\section{Introduction}

\IEEEPARstart{W}{ith} the introduction of the concept of virtual factories and intelligent manufacturing, the reliability of cross-organizational data sharing has received greater attention\cite{singh2020blockiotintelligence}. As a valuable data source, the IoT provides the resources that are required for manufacturing processes, intermediate goods, and manufactured goods. These resources are often enriched with sensor, identification, and communication technologies, such as RFID tags\cite{schulte2012plug}. Although many IoT architectures have been generated over the past two decades, they still have problems with trust, security, overhead, and scalability\cite{cao2019internet}. 

\begin{figure}[htbp]
    \centerline{\includegraphics[width=3.25in,trim=25 18 25 20, clip]{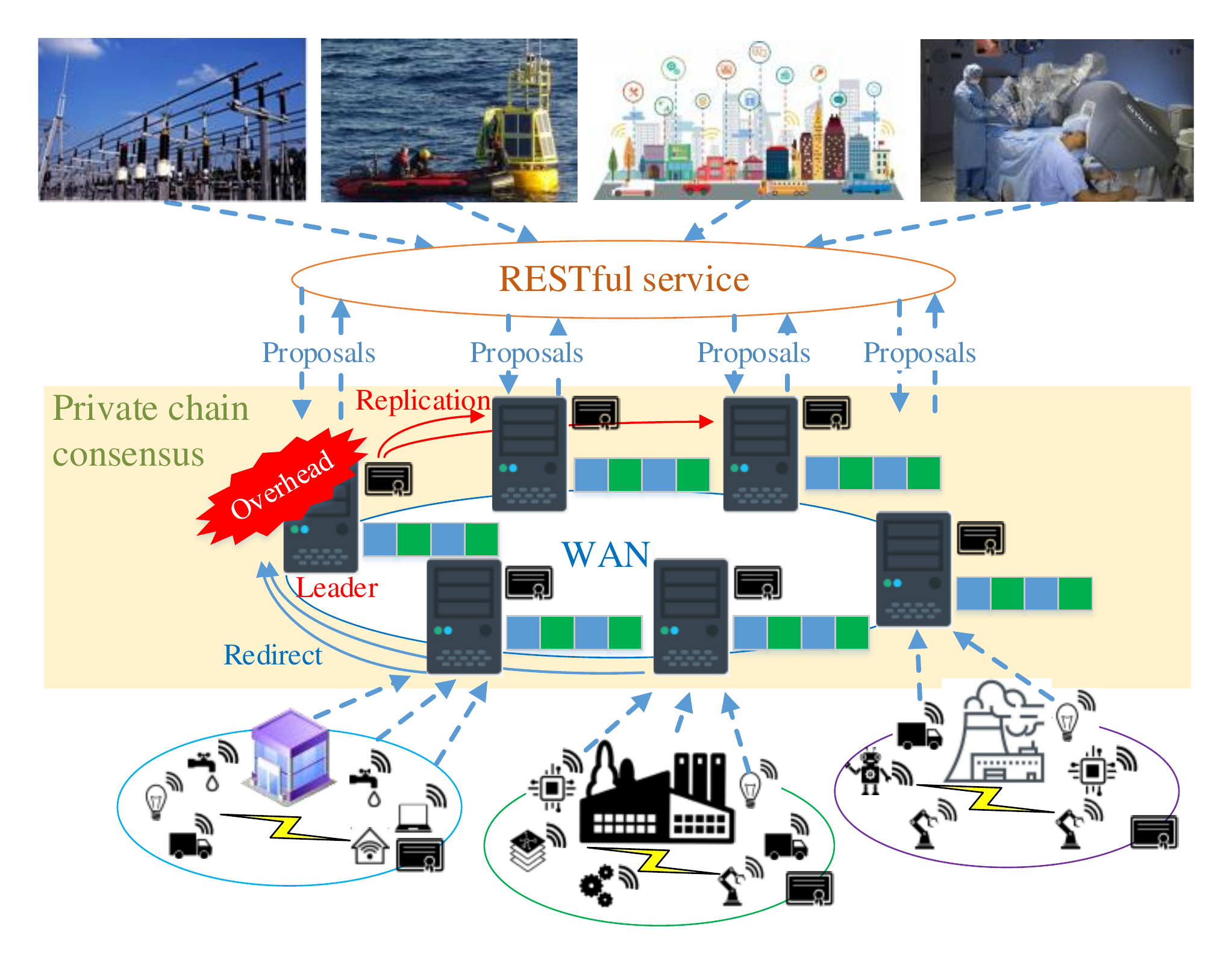}}
    \caption{The architecture of permissioned private chain. In this paper, we focus on the optimization of the consistency model to provide better performance and scalability by alleviating the overhead on the leader.}
    \label{fig001}
\end{figure}

The public blockchain architecture based on proof of stake (PoS)\cite{kiayias2017ouroboros} and proof of work (PoW) \cite{nakamoto2008peer} solves trust and security issues well. With good resistance to malicious behaviors, its application in virtual currency has been widely recognized. However, the throughput (lower than 30 TPS) and confirmation delay (minutes) that they provide cannot meet the demands of IoT systems\cite{bandara2018mystiko,cao2020performance}. To address the problems, direct acyclic graph (DAG)-based approches represented by Tangle \cite{popov2018tangle} improve the micro-transactions processing performance well. Its confirmation delay depends on the confirmation threshold $\alpha$ and the dependencies of transactions\cite{li2020direct}. To build trust between organizations while ensuring high transaction processing efficiency, permissioned private chains use leader-based Crash Fault Tolerant (CFT) consensus protocols (such as Paxos\cite{lamport2001paxos}, Raft\cite{ongaro2014search}) and reinforced identification checks\cite{chen2019full} to achieve state-machine replication (SMR). Different from Byzantine Fault Tolerance (BFT) protocols (such as PBFT\cite{castro1999practical}, PoS\cite{kiayias2017ouroboros}, PoW\cite{nakamoto2008peer}), CFT does not allow the existence of Byzantine nodes in the network. Nonetheless, CFT protocols are designed for low-latency distributed systems\cite{huang2019performance,corbett2013spanner,aguilera2020microsecond}. For IoT blockchain systems, sensor data are transmitted using (wireless) sensor networks and are provided through well-defined interfaces, and consensus servers are deployed in Wide Area Network (WAN) environments where the network latency between nodes is unstable\cite{zhou2017reinforcement,hasenburg2020geobroker,cao2019internet}, as shown in Fig.~\ref{fig001}.

As a result, this limitation has several important consequences. First, the massive IoT data lead to data retransmissions when performing SMR due to the high latency, which generates high network traffic between consensus nodes. Second, the centralized SMR incurs worker thread starvation problems on the leader node, which causes a significant decrease in the TPS performance of the blockchain system and increases the high overhead on the leader. Third, followers need to redirect the data to the leader for consistency, which generates additional latency for communicating with a remote leader.

\begin{figure}[htbp]
\centerline{\includegraphics[width=3.5in,trim=15 15 15 15, clip]{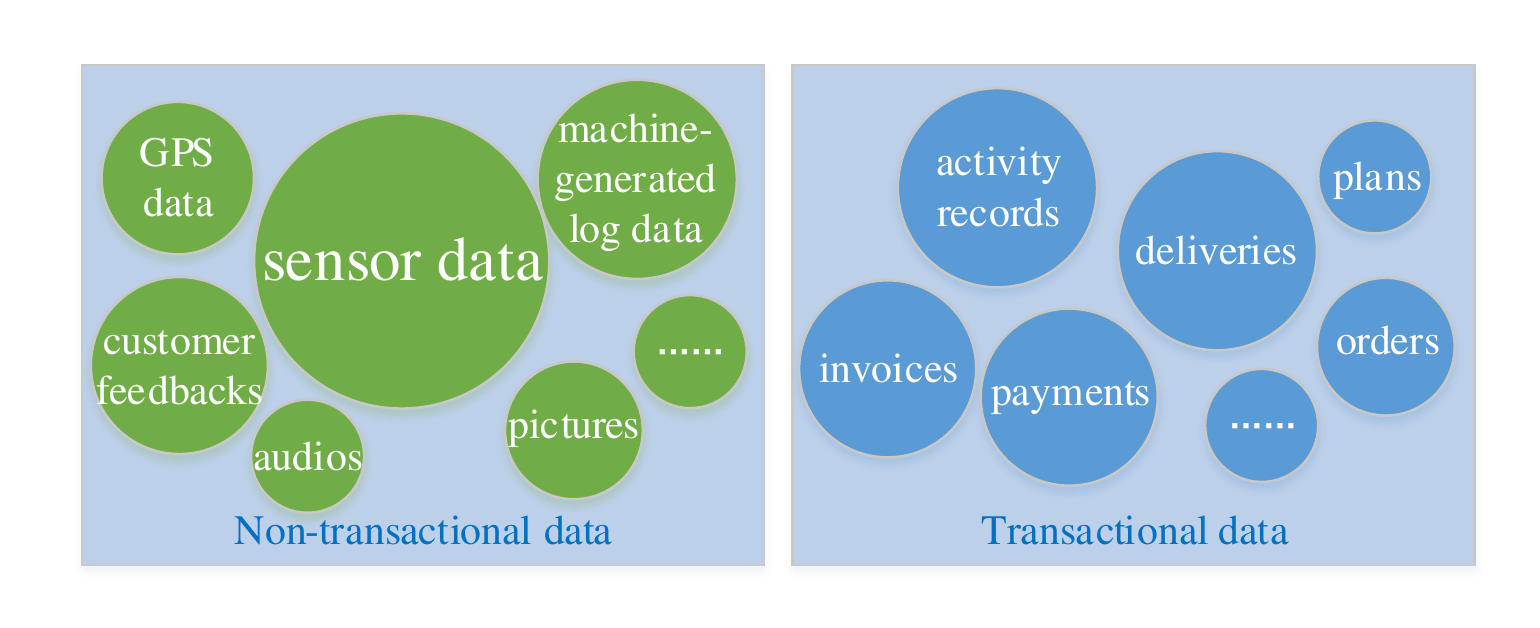}}
\caption{Examples of transactional and non-transactional data in IoT systems.}
\label{fig002}
\end{figure}

Through research on the data generated by the IoT system, we found that operational non-transactional data are generated at a scale. Different from transactional data, these data contain up-to-date, real-time status information and are less likely to be updated\cite{li2012storage}\cite{chandra2015base}\cite{viotti2016consistency}. Therefore, the availability and timeliness of non-transactional data are the focus of researchers\cite{terry1994session}\cite{saito2005optimistic}\cite{vogels2009eventually}. To more clearly distinguish the difference between transactional data and non-transactional data, we give an example in Fig.~\ref{fig002}.

Unfortunately, the existing CFT-based IoT blockchain system does not have optimized solutions for non-transactional data, i.e, transactional data and non-transactional data are treated the same. Most existing optimization methods for the geo-distributed consistency are implemented with multi-leader, i.e., the data store is divided into shards with different leaders. Although it can significantly reduce the load of the single leader under high latency, when any node is down, the corresponding shard will be temporarily unavailable due to the re-election. Another type of method dynamically designates leaders\cite{moraru2013there}, but its performance is sensitive to the degree of dependence in the data and is likely to cause inconsistency when the number of nodes is large\cite{zhao2018sdpaxos}.

We want to reduce the load generated by SMR on the leader in the geo-distributed scenario without damaging the power of the leader and changing the election process, thereby improving the performance of the consistency model. It is common to implement SMR using replicated logs. Therefore, our goal is to make the followers undertake as much non-transactional log replication work as possible. The key point is how the logs replicated by the followers are confirmed by each node to reach a consensus. Among them, it is a challenge to design a novel control method for the leader in the log replication process, which can solve the log replication failure problem caused by the downtime of the follower nodes\cite{ganesan2020strong,ahn2019designing}. Besides, how to avoid log submission conflicts when followers cannot obtain the entire cluster status in time is also an important problem that we need to solve.

In this paper, we propose Leader Confirm Replication (LCR), a method for optimizing leader-based consensus performance in geo-distributed consensus. LCR is the first SMR optimization method that does not change the number of leaders and election strategy and does not add strong control roles to non-leader nodes. It achieves consistency by proposing the future-log, which makes followers become the data-leader to replicate the non-transactional data without conflict. The leader only needs to confirm the right replicated future-log by a signal to achieve quorum. In doing so, LCR solves two major problems that we detailed above in geo-distributed scenarios: 1) It decouples the non-transactional data replication process from normal SMR, thereby reducing the leader's resource consumption and improving the consistency efficiency of the data. 2) The entry status information included in each SMR request is increased by applying signal entries and a leader confirmation mechanism. The contributions include the following points:

1) We analyzed the cause of the concentrated load problem in the leader-based consistency model and discussed the existing optimization models.

2) We designed LCR, detailed the future-log replication model, realized nonconfliction log replication, and proved its availability. In addition, we designed generation to resolve log conflicts that can be caused by membership changes in the cluster.

3) We implemented LCR based on Raft and compared its performance with popular approaches. Then, we verified its availability and the optimization effect on the request latency, TPS, network traffic, and CPU load in ms-latency network environments.

4) We analyzed the overhead of LCR and discussed the application scenarios, which can provide ideas for follow-up studies.

We arranged this paper as follows: In Section II, we detail the related work, and in Section III, we show the problems and motivation for the research. We introduce the future-log replication method of LCR in Section IV, and we explain and prove the availability of the model in Section V. In Section VI, we present the experiments that were conducted and the analysis on LCR, and in Section VII, we discuss the protocol. We summarize our work in Section VIII.

\section{Related work}

\subsection{Consensus protocols in private blockchains}

The rapid development of private chain technology represented by Hyperledger has made geo-distributed blockchain systems widely used\cite{muzammal2019renovating}. Different from the public chain, any node needs to go through a reinforced identification check before it can participate in the blockchain network\cite{chen2019full}. Therefore, malicious behavior in the private chain will be easily detected and traced. It allows the private chain system to weaken the security design in an anonymous environment and adopts a consensus model that supports better TPS performance and scalability\cite{vukolic2015quest}. Therefore, the existing widely used private chain architectures use CFT protocols (e.g., Raft\cite{ongaro2014search}, Paxos\cite{lamport2001paxos}) instead of BFT protocols (e.g., PBFT, PoW, DAG), thereby improving the cluster scalability and the efficiency of the consistency\cite{huang2019performance,mingxiao2017review}.

\begin{table*}
    \renewcommand{\arraystretch}{1.1}
    \caption{Performance comparison of LCR and state of the art consensus models}
    \label{table_0}
    \centering
      \setlength{\tabcolsep}{1.2mm}{ 
      \begin{tabular}{llllll}
      \toprule
      Consistency  & Byzantine Fault  & Single node  & Leader strategy & Client response & Major drawback\\
      model&Tolerate&overhead&&&\\
      \midrule
      PoW\cite{nakamoto2008peer} & Yes & High & No leader & Tens minutes &High overhead and long confirmation delay\\
      PoS\cite{kiayias2017ouroboros}   & Yes & High & No leader & Tens seconds &Long confirmation delay\\
      DAG\cite{popov2018tangle} & Yes & High & No leader & Subsecond & High dependence on data business\\
      PBFT\cite{castro1999practical} & Yes & High & Centralized & 3 RTT&Low scalability\\
      Raft\cite{ongaro2014search}	& No & High & Centralized & 2 RTT &Single point overhead\\
      Multi-Raft\cite{huang2020tidb} & No & Medium & Centralized & 2 RTT&Frequent shard leader re-election\\
      Mencius\cite{barcelona2008mencius} & No& Medium & Rotational & 2 RTT & Run at slowest node\\
      Multi-Paxos\cite{lamport2019part}& No& High & Centralized &2 RTT&Single point overhead\\
      EPaxos \cite{moraru2013there} & No& Low & No leader & $\geq$ 1 RTT&Higher CPU consumption and more round trips \\ 
      &&&&&for analyzing dependency\\
      SDPaxos \cite{zhao2018sdpaxos} & No& Medium & Centralized &1.5 RTT &More connections between nodes\\ 
      LCR & No& Low&Centralized&2 RTT (transactional) &Our solution\\ 
       & & && $\geq$ 1 RTT (non-transactional)&\\ 
      \bottomrule
\end{tabular}}
\end{table*}

However, CFT protocols are designed for low latency environments. In high network latency, most geo-distributed CFT protocols use leader-based SMR to achieve consistency, which can avoid the performance loss caused by replication conflict in the SMR process (e.g., the livelock in Paxos\cite{michael2019teaching}). Nonetheless, the overhead of the leader will cause serious performance degradation when the consistency model runs in high concurrency. Although it is a good idea theoretically to dynamically change the controller of the consensus process to the follower nodes, it requires a stable network environment between each node to obtain the real-time node status. Otherwise, the consistency model will run at the speed of the slowest node\cite{moraru2013there}. 

\subsection{Optimization and application of CFT protocols in private blockchain}

As we discussed in Section II.A, the CFT consistency model which is commonly used in private chains, meets leader overhead in geo-distributed environments. To address these issues, existing popular approaches reduce the leader overhead by adopting multiple leaders or reducing the frequency of the log replication times by batching operations. We summarize the consensus models mentioned in this section in TABLE 1.

\subsubsection{Leader or leaderless}

Leaderless consistency protocols represented by Paxos are widely recognized. Researchers have optimized Paxos and proposed some variants to improve its performance in high-latency environments. Since Paxos\cite{lamport2001paxos} could have livelock problems in WAN environments, multi-Paxos adds the leader role to it. However, this strategy also places a higher load on the leader node. Mencius\cite{barcelona2008mencius} makes each node take turns to act as the leader to solve this problem, but this approach makes the performance be determined by the slowest node. Therefore, EPaxos\cite{moraru2013there} maintains a leaderless design, which determines the committing method by analyzing the dependencies between operations. If there is no dependency between the operations, then it will use the fast path to reach an agreement. If there is a dependency conflict, an additional round trip is needed to execute ordering constraints to ensure causal consistency. To reduce the extra delay caused by dependency conflicts, SDPaxos\cite{zhao2018sdpaxos} adds a sequencer (which can be understood as the leader) to order the data with dependency. Although it reduces the round trip by 0.5 ,it comes at the cost of a heavier load on the leader and more requests between nodes, which are caused by O-instance. Raft is a simplified version of Paxos. Its single leader strategy requires only one round trip to complete the operation submission. The simplicity of its implementation also makes it widely used in industry and blockchain systems. However, its single-node load problem is an important factor that affects its performance.

\subsubsection{Multiple leaders/controllers}

To avoid conflicts of control rights, the multi-leader model divides the data store into multiple shards with different leaders\cite{yang2019blockchain,howard2015raft}. This method distributes the log replication operation originally led by one node to each node for execution. Although this approach can reduce the overhead generated by the SMR model, making each node act as a leader will make the data replication requests between nodes more frequent. Therefore, TiDB\cite{huang2020tidb} uses connection reuse to ensure that the replication request and response include the metadata of logs and entries, thereby reducing the number of requests between nodes. However, it still cannot solve the frequent leader re-election problem. When any node in the multi-leader cluster is down, the cluster must invoke re-election; otherwise, the shards that it controls will be unavailable because the node must be the leader of a shard. In addition, during the leader election process, no consensus can be established in the corresponding shards, which will cause availability issues. Therefore, multi-leader strategy is more suitable for use in a low-latency, high-reliability environment.

\subsubsection{Batching entries}

In addition to multi-leader, reducing the number of requests in the SMR protocol is another popular approach to improving the TPS. When the new entries are committed to the leader, using the pipeline to replicate them to the followers is a widely used implementation method of the SMR protocol\cite{zhang2020dependency,carmeli2004high,friedman2006adaptive}. Therefore, researchers designed the buffer to merge the entries and replicate them in one request, which can significantly reduce the SMR requests between nodes, thereby achieving the effect of improving the TPS\cite{cao2018polarfs,wang2017fast}.

However, this approach has two potential problems. 1) The average response time will increase. Batching makes the first entry in the buffer wait for subsequent entries to finish writing until the buffer is full, after which the replication method can be invoked by the leader. Nevertheless, it can significantly reduce the replication requests, thereby reducing the load of the leader in high concurrency scenarios. However, when the load is low, the buffer requires a longer time to be fully filled, which will increase the response time\cite{lamport2019part}. 2) The followers must keep more connections at the same time. The reason is that the leader must wait for the entries to be applied before returning the success message to the client. Therefore, the followers that forward the client request to the leader must also keep the requests alive. However, this approach will increase the latency, especially in single-ms or higher network latency environments.

\section{Problem statement}

The process of SMR between nodes is sensitive to network latency, which easily makes nodes prone to performance problems in geo-distributed systems. Therefore, in this section, we analyze the SMR process in a geo-distributed environment.

\subsection{SMR performance with million-second latency}

Raft numbers the entries and copies them to the follower nodes in order\cite{sharma2020blockchain}. Thus, Raft can use the $lastIndex$ of each node to quickly reach a consensus. However, private chains are widely used in supervised multi-subject data-sharing scenarios. The entries they generate contain only a single database operation on new keys. In IoT systems, the leader performs log replication more frequently since the operational non-transactional data are generated on a scale. In addition, the private chain systems need to be deployed in each organization’s datacenter to ensure that they can participate in the consensus\cite{yao2019bla}. However, we found that the network traffic and I/O overhead generated by the SMR model are very sensitive to network latency. The overhead on the leader will incur higher network overhead and TPS degradation. In addition, most non-transactional requests only insert new key values. The operation of verifying the input and output of the transactions will increase the leader's computing load, resulting in a significant increase in the response time, and adversely affecting the QoS of the private chain system\cite{zhang2020byzantine}.

To better understand the impact of ms-level latency on the efficiency of consistency in geo-distributed systems, we performed experiments on a real-time cross-datacenter backup system for Kingsoft Cloud's database, which runs with a 1.5-2 ms network latency, and a Hyperledger-based private chain system, which runs with a 3-7 ms latency. The purpose is to obtain the performance bottleneck of geo-distributed SMR that enables us to make realistic assumptions for the design of LCR-Raft.

\begin{figure}[htbp]
    \centerline{\includegraphics[width=3.5in,trim=0 0 526 305, clip]{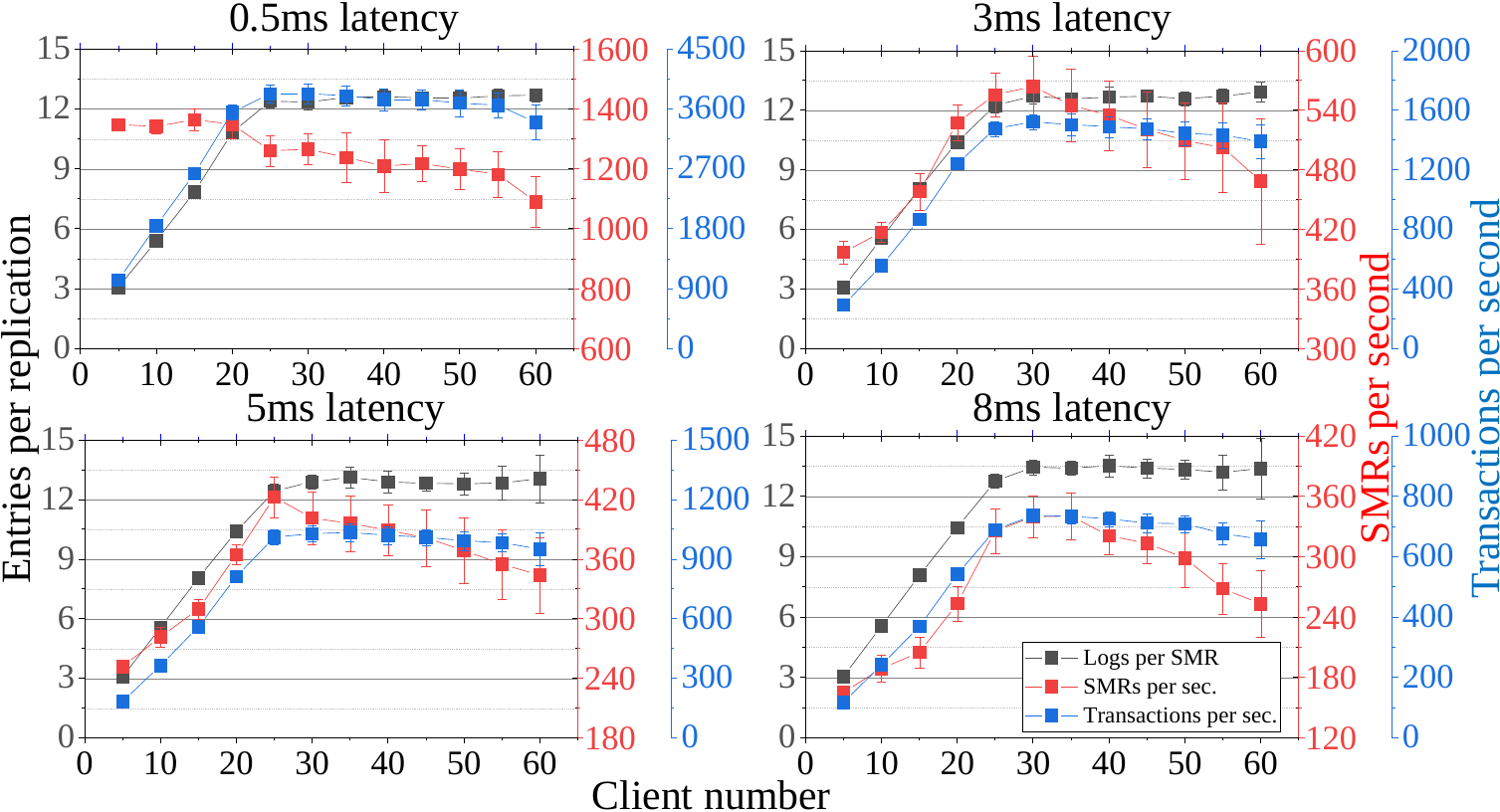}}
    \caption{Different network latencies are added to the 5-node Raft cluster. The x-axis is the number of clients, and the latency fluctuation range and occurrence are 0.1 ms and 30\%, respectively. Each request from the client is regarded as a transaction.}
    \label{fig003}
    \end{figure}

In experiments, we send requests concurrently to Raft with numbers of users, and count its TPS and entries' retransmission times under different network latencies, as shown in Fig.~\ref{fig003}.

\begin{figure}[htbp]
\centerline{\includegraphics[width=2.2in,trim=0 0 646 386, clip]{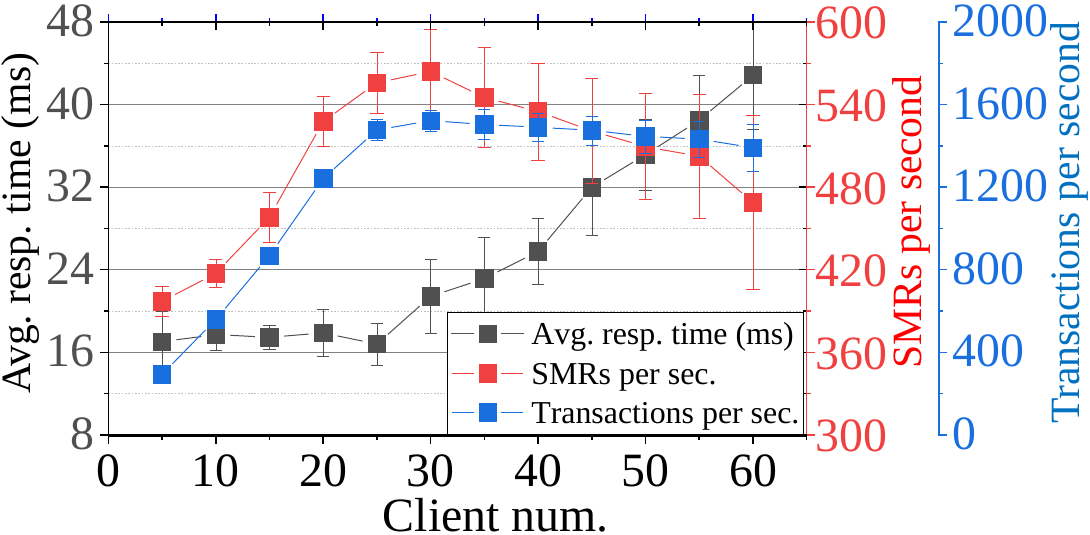}}
\caption{We conducted a response time experiment under the environment of 3 ms network latency and 5 nodes. When the number of clients is greater than 25, the average response time gradually increases, and the TPS and replications per sec decrease.}
\label{fig004}
\end{figure}

As the number of clients increases, the TPS and the entries per replication of the cluster gradually rise and reach peak performance when the number of clients is less than 25. At this stage, the cluster does not reach its transaction processing capacity, and thus, a stable response time can be guaranteed. With the increase in the number of clients, we observed that the frequency of replication was significantly reduced and the response time increased, as shown in Fig.~\ref{fig004}. This outcome occurs because the leader backlogs clients requests or SMR requests due to high network latency, which leads to worker thread starvation or even deadlocks on the leader\cite{moniz2017blotter}. Therefore, the limited processing capacity of the leader is a key performance bottleneck of leader-based consistency models in geo-distributed scenarios.

\subsection{Network traffic feature of leader}

\begin{figure}[htbp]
\centerline{\includegraphics[width=3.3in,trim=0 0 440 211, clip]{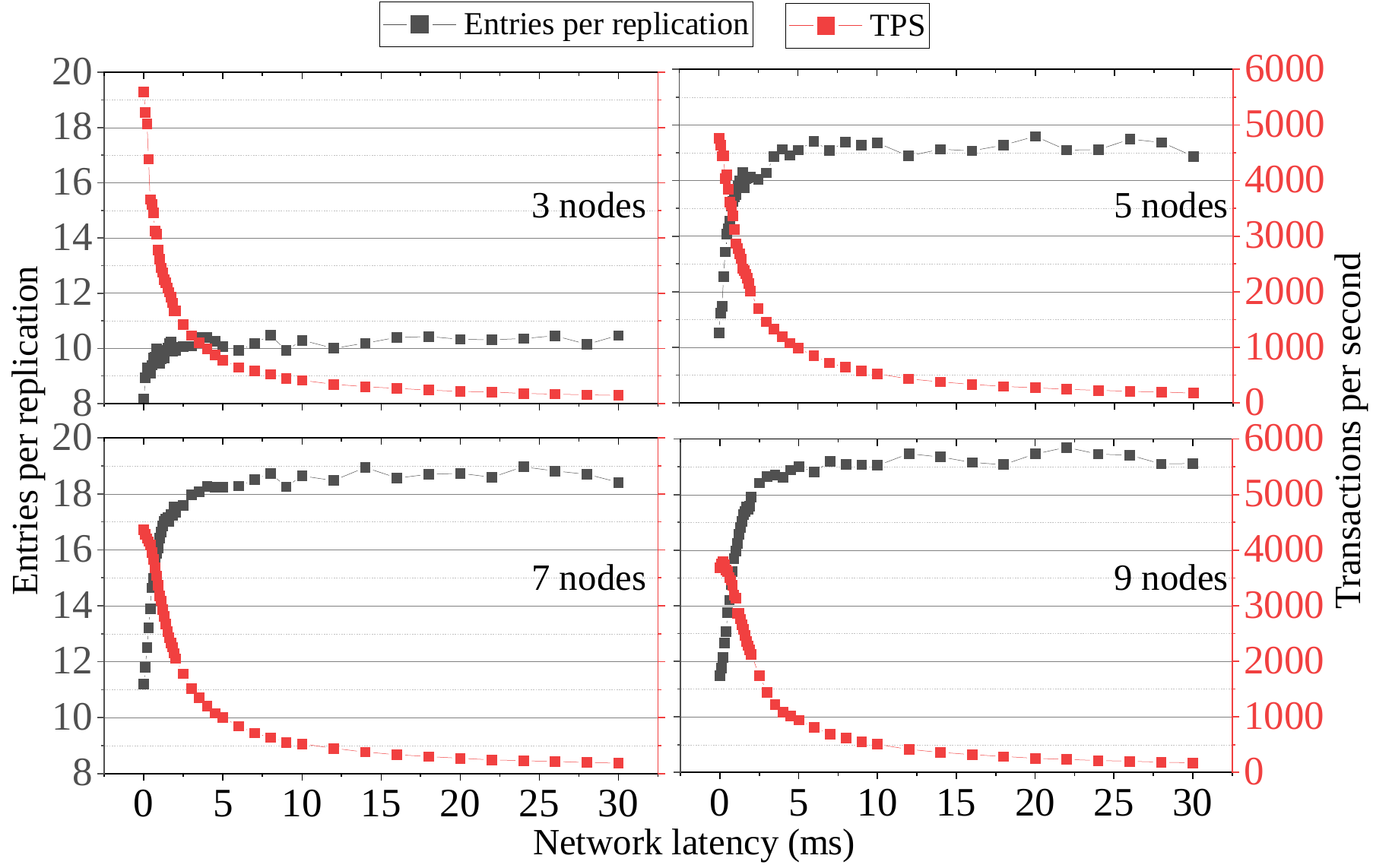}}
\caption{In the experiment, we added different network latencies to the Raft cluster with 3, 5, 7, and 9 nodes and used a pipeline strategy to replicate entries. The x-axis is the average network latency, and the latency fluctuation range and probability are 0.1 ms and 30\%, respectively. It can be seen that as the network latency increases, the entries per replication will increase by 60\%-70\%}
\label{fig005}
\end{figure}

To express the network load feature of the SMR model more clearly, we took Raft as an example to analyze its log replication process. Ongaro provides the pipeline and batching approaches to replicate entries in the initial Raft paper\cite{ongaro2014search,ongaro2014consensus}. In practice, there are three main ways to implement it: serial-batching\cite{SOFA-JRaft}, pipeline\cite{etcd,huang2020tidb}, and full parallel\cite{raftjava}. For LAN environments, high routing performance and a stable network are provided. Therefore, it is a better choice to use full parallel or pipeline methods since it can provide a lower response time. However, in geo-distributed systems, network resource are more expensive. In addition, the unstable network environment also makes designers tend to adopt a conservative replication strategy. Accordingly, pipeline strategies were carefully designed by \cite{huang2020tidb} to improve the efficiency of the log replication in geo-distributed scenarios. In an ideal situation, each entry is transmitted k(n-1) times by the leader. However, estimating the round-trip time (RTT) is a key problem for pipeline-based approaches\cite{ongaro2014consensus} and is not an easy task in WAN environments\cite{yasuda2018prediction}. For verification, we counted the entry retransmission of the pipeline-based Raft under different network latencies in Fig.~\ref{fig005}.

Therefore, it is one of our research motivations to reduce the leader's message transmission cost in the WAN environment.

\section{Log replication in LCR}

LCR designed the future-log to distribute the leader's log replication work to followers to alleviate its concentrated load in the consistency model. The future-log can be led by any follower and applied to the state machine using a leader confirmation mechanism. Similar to the log in Raft (we call it normal-log in this paper to distinguish it from future-log), future-log is saved as a segmented log on disk and consists of entries and metadata. We call the entries in the future-log future entries and entries in the normal-log normal entries. To make it easier for readers to understand, the figures in this paper combine future-log and normal-log, because future entry and normal entry will not conflict on the index in the same node. 

\begin{figure}[htbp]
    \centerline{\includegraphics[width=3.8in,trim=0 20 0 20, clip]{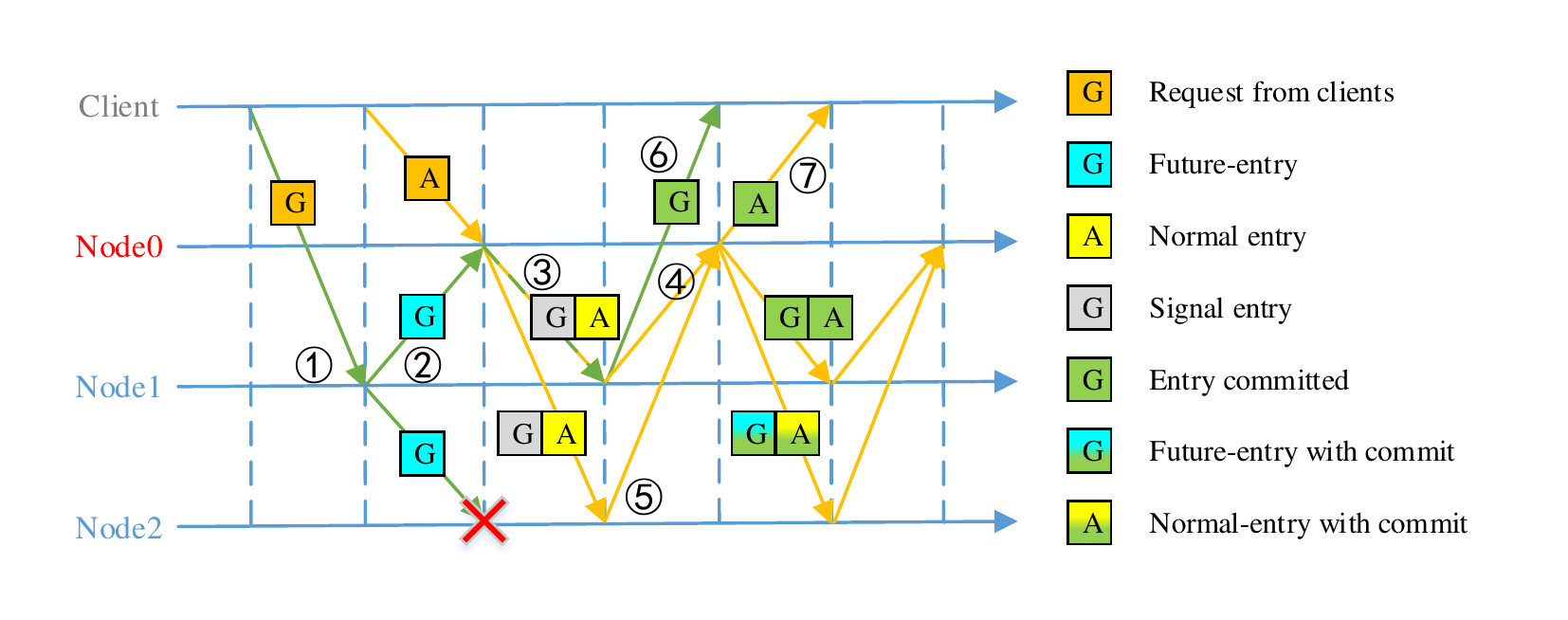}}
    \caption{The data-leader will return to the client a success message when it has copied the future-log to most nodes that contain the leader. The future-log will then be confirmed by the leader with a signal and copied to the followers who did not receive the future-log.}
    \label{fig006}
    \end{figure}

Therefore, LCR will not weaken the power of the leader in the replication process. In addition, since the leader uses the confirmation signal to replace the detailed content of the entry, it reduces the leader's network resource consumption. Besides, LCR guarantees that followers have the same standing, which greatly reduces the complexity of the protocol design and makes the membership change process clearer than than multi-leader types of approaches. Here, we detail the leader confirmation mechanism with the messaging model of LCR in Fig.~\ref{fig006} and TABLE~\ref{table_1}.

\begin{table}[htbp]
  \scriptsize
  \renewcommand{\arraystretch}{1.3}
  \centering
    \caption{Detailed steps of the leader confirmation mechanism in Fig~.\ref{fig006}}\label{table_1}
    \begin{tabular}{p{0.2in}p{2.8in}}
    \hline
    Steps &Operations\\
    \hline
    1& The follower Node 2 receives the request from the client. Node 2 will create a future entry G, assign a future index to it, and write it to the future log. \\
    2& Node 2 replicates G to other nodes. \\    
    3& The leader receives a request from the client, it will create a normal entry A and assign it a normal index. Subsequently, the leader found that entry G in the future-log has not been applied to the normal log, the leader assigns a normal index to it and writes it to the normal log. Then, the leader appends normal entries A and signal entry G to each follower. \\    
    4& Node 1 receives the Signal log G, takes the corresponding future entry from the future log according to its future index, replaces the signal log with it, and applies G and A to the normal log. When it is completed, Node 1 responds to the leader that it has applied the index of entry A. \\    
    5& Node 2 receives the signal log, but since it cannot find the future entry G in the future log, it will return index-1 of the signal entry G as its lastAppliedIndex to the leader. \\    
    6& Node 1 knows that future entry G has been committed, and it will return a success message to the client.\\
    7& The leader learns that more than half of the nodes have applied A; it will commit the entry whose normal index is smaller than A and notify all of the other nodes to commit these entries. Since Node 2 did not successfully apply future entry G, it transmits the raw data of G to Node 2 along with the committed index. \\    
    \hline
    \end{tabular}
    \end{table}
    
\begin{figure}[htbp]
        \centerline{\includegraphics[width=3.8in,trim=0 15 0 20, clip]{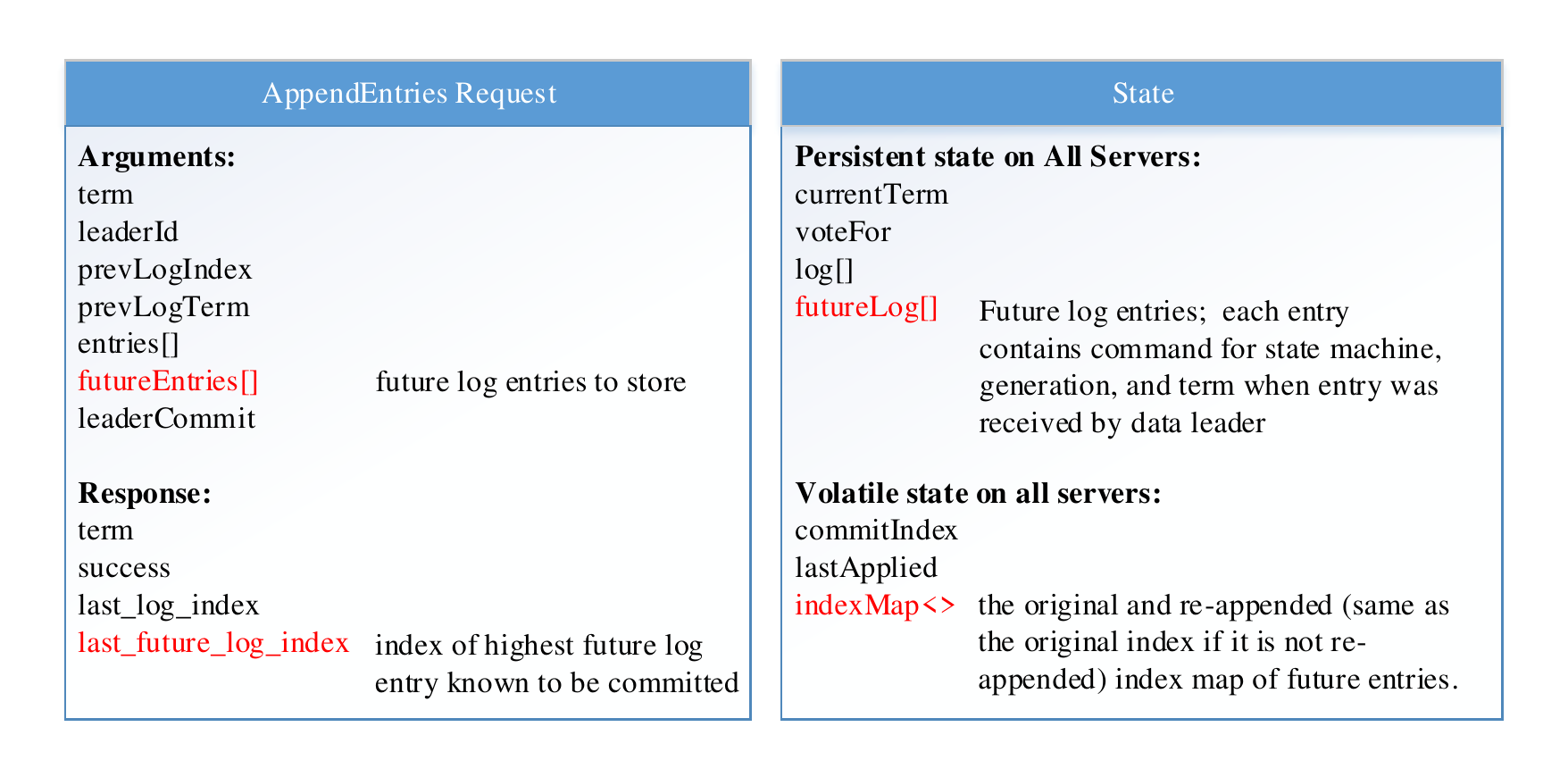}}
        \caption{LCR protocol design. The red attributes are additional in LCR}
        \label{fig007}
        \end{figure}

    Besides, LCR does not change the leader election attributes. Taking Raft as an example, LCR only needs to add the $futureEntries$ property to the request of $AppendEntries$ to replicate the future-log between nodes. In the response of $AppendEntries$, the $lastIndex$ of the future-log attribute is added, which is used to acknowledge the data-leader the max index of the future-log that it has applied, as shown in Fig.~\ref{fig007}.

    Therefore, LCR is still stateless in the future-log replication process, which makes it possible to maintain a low coupling between the nodes and tolerate unstable network environments.

In this section, we answer the following questions: 1) how the future log is generated and replicated to all nodes; 2) how to resolve the conflict when normal log and future log are allocated the same index; and 3) how to reduce the occurrence of conflicts. TABLE~\ref{table_2} collects our notation.

\begin{table}[htbp]
\scriptsize
\renewcommand{\arraystretch}{1.3}
\centering
\caption{Notations used in this paper}\label{table_2}
\begin{tabular}{p{0.5in}p{2.5in}}
\hline
Name & Meaning\\
\hline
$S_i$ & a sequenced number that identifies a node.\\
$lastIndex_i$ & the max index of entry that was saved in the segmented log of future-log or normal-log on node $S_i$. \\
$nextIndex_i$ & the next index that need to be replicated to $S_i$. \\
$D_{t}$& a transactional data that is generated from a client. \\
$D_{nt}$& a non-transactional data that is generated from a client. \\
$G_i$& the current $generation$ of node $S_i$. \\
$E_\lambda$& normal entry with index $\lambda$.\\
$FE_\lambda$& future entry with index $\lambda$.\\
$\Delta_c$ & the network latency between the cluster and client.\\
$\Delta_n$ & the network latency between the consensus nodes in a cluster.\\
$\eta_l$ & the time period from normal entry packaging to replication request sending on the leader.\\
$\eta_f$ & the time period from future entry packaging to replication request sending on followers.\\
$\eta_w$ & the time period that a data-leader waits for a signal entry to apply data\\
$\iota$ & the latency that is composed of disk operations latency such as log writing and state machine applying.\\
\hline
\end{tabular}
\end{table}

\subsection{Future-log replication}

The numbering method of future entry is similar to the commonly used creasing numbering in the SMR protocol, to make it more applicable to data consistency models. The difference is that the index of two entries that are continuously appended to the future-log can be discontinuous. In addition, the replication process of future-log is led by followers. We call the follower leading a certain future entry replication a data-leader and other nodes are data-followers. When a follower with id $S_i$ receives non-transactional data $D_{nt}$, the follower will become the data-leader, allocate a future log index $o$ to $D_{nt}$ and assemble it into a new future entry $FE_{\lambda}$. The index allocation method is shown in Equation (1). 

\begin{equation}
    \lambda = S_i + G_i + lastIndex_i - lastIndex_i \% G_i
\end{equation}
where $lastIndex_i$ is the max index of future entries that $S_i$ saved, and $G_i$ is the current $generation$ of $S_i$ ($generation$ is an increasing integer as the number of nodes changes and $generation >= max(S_i)$, which we introduce in detail in Section V).

\textit{Theorem 1.} The index of future entry $FE_{\lambda}$ generated by $S_i$ is globally unique.

\textit{Proof. } For $FE_{\lambda}$, we can prove that 

\begin{equation}
    \lambda \% G_i = S_i
\end{equation}
because 
\begin{equation}
G_i \% G_i = 0
\end{equation}
,
\begin{equation}
(lastIndex_i - lastIndex_i \% G_i) \% G_i = 0
\end{equation}
and 
\begin{equation}
G_i > max(S_i)
\end{equation}
Therefore, for servers $S_i$ and $S_j$, $G_i = G_j$ , and the index of future entry that $S_i$ generates must be different from another server $S_j$.

When $G_i > G_j$, $S_i$ will reject the copy request from $S_j$. For $S_j$ with an obsoleted $G_j$, when $S_j$ receives a message from a node with a newer $generation$ $G_i$, $S_j$ can find all of the future entries that it created according to Equation (1). Then, $S_j$ reallocates a nonconflicting index for these future entries with $G_i$ (we will detail how servers address the $generation$ change in Section V.)

\begin{figure}[htbp]
    \centerline{\includegraphics[width=3.5in,trim=0 15 0 15, clip]{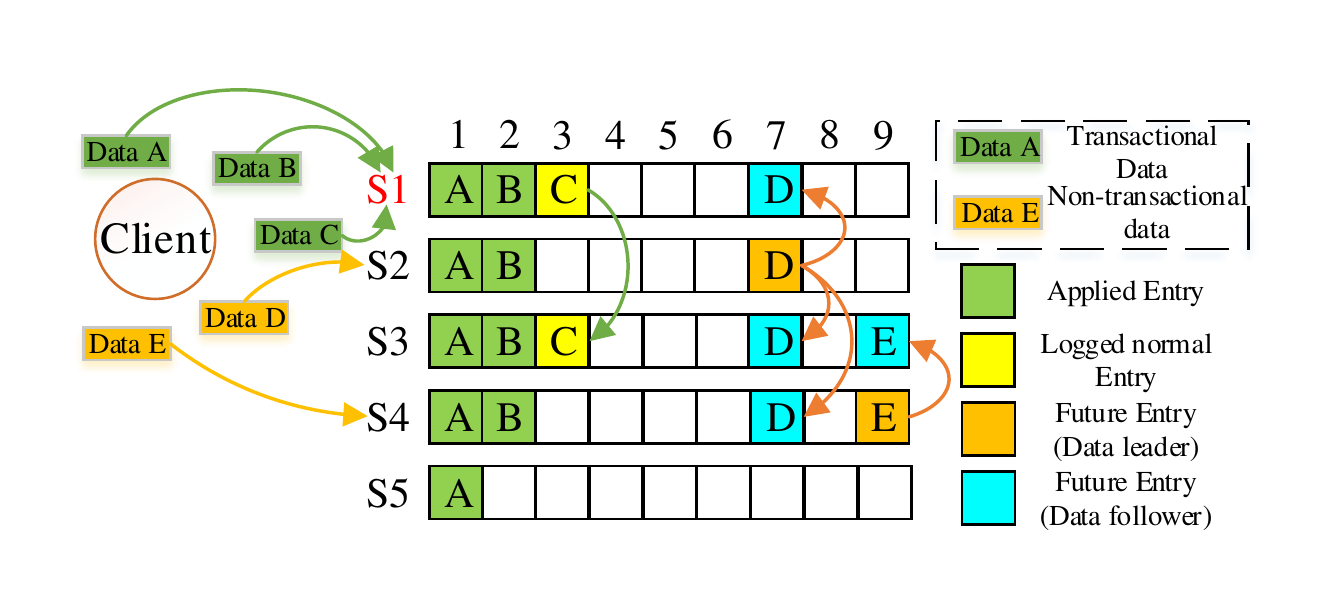}}
    \caption{Example of the future entries replication process, where S1 is the leader}
    \label{fig008}
    \end{figure}

After $FE_{\lambda}$ is written on a future-log and committed by the data-leader, it copies $FE_{\lambda}$ to all nodes, as shown in Fig.~\ref{fig008}.

The node will reject the future entry replication request from the nodes with a smaller $term$ or $generation$ and return the newer information that it knows to them. It ensures that nodes can reach a consensus on the same future entries if and only if the nodes have the same $term$ and $generation$. The future entries generated from a node with an obsoleted $term$ or $generation$ will not be applied because they cannot be accepted by quorum. It is very important because the signal entry must correspond to the same future entry on all nodes.

\subsection{Signal entry and re-replication}

However, the data-leader's log replication process could have two problems: 1) The leader receives future entry $FE_\lambda$, but parts of the nodes do not receive the $FE_\lambda$. 2) The leader did not receive the $FE_\lambda$ from the data-leader. 

\begin{figure}[htbp]
    \centerline{\includegraphics[width=2.8in,trim=0 15 0 20, clip]{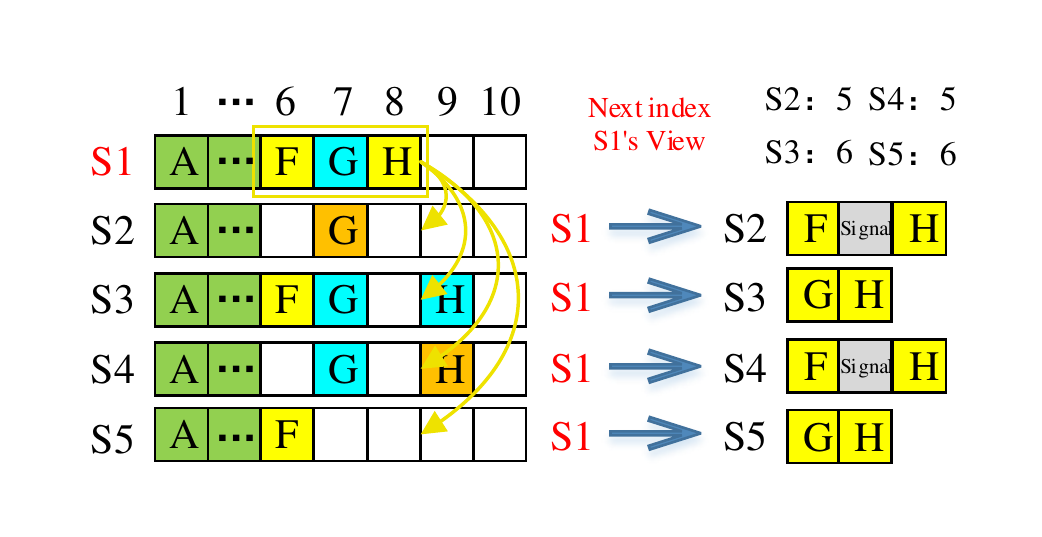}}
    \caption{Example of the signal entry and re-replication process. Since the $nextIndex$ of S3 and S5 is 6 and the entry with index 7 is the future entry, S1 believes that S3 and S5 have not received $FE_7$ (even though S3 has received it).}
    \label{fig009}
    \end{figure}

For the first case, leader $S_i$ does not know which followers have received the $FE_\lambda$. Therefore, the leader will use signal entry to replace $FE_\lambda$. In particular, if the $nextIndex_j+1$ of follower $S_j$ is a future entry, $FE_j$ will not be replaced by signal entry. As shown in Fig.~\ref{fig009}, S5 did not obtain future entry with $FE_7$, and thus, it can only commit a normal entry with $E_6$ and set $lastIndex_5$ to 6 in the return value. For a follower $S_k$  whose $nextIndex_k+1$ indexed entry is a future entry, the leader believes that it has not received the future entry $FE_{nextIndex_k+1}$. Therefore, when the leader is packaging entries for replication, it will assemble the entry as a normal entry. Followers will trust the correctness of the leader and commit the entries.
 
For the second case, i.e., when follower $S_i$ with $G_i$ finds that the normal entry $E_\lambda$ sent by the leader conflicts with $FE_\lambda$, the follower trusts the leader. Then, it will check whether the data-leader of $FE_\lambda$ is itself according to Equation (6).

\begin{equation}
    \lambda \% G_i == S_i
\end{equation}

\begin{figure}[htbp]
    \centerline{\includegraphics[width=2in,trim=0 15 0 20, clip]{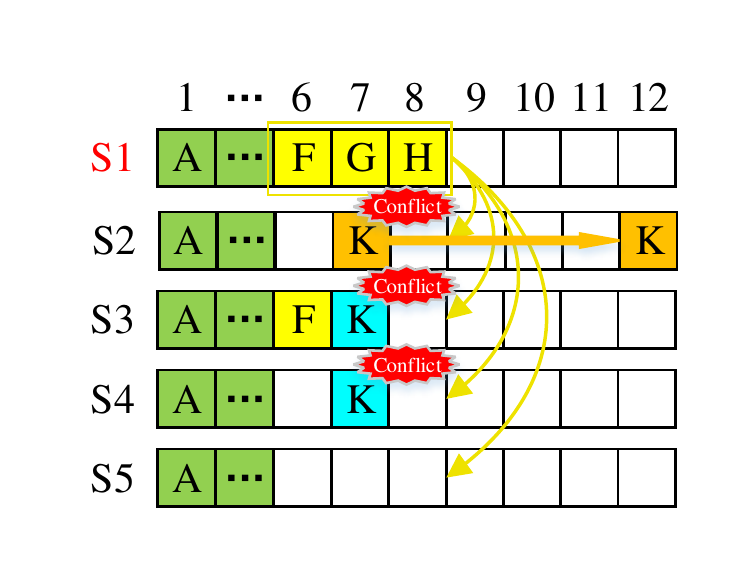}}
    \caption{Example of an entry conflict: each node will unconditionally trust the leader. When a conflict occurs, all nodes except the data-leader will discard $FE_7$, and $FE_7$ will be reassigned to a new index for the next replication of S2.}
    \label{fig010}
    \end{figure}

If the result is $true$, then, $S_i$ is the data-leader of $FE_\lambda$, and $S_i$ deletes $FE_\lambda$ from the future-log, reallocates the future entry index according to Equation (1), and rewrite it into the future-log. If it is $false$, it will discard $FE_\lambda$. Therefore, regardless of whether the result is $true$ or $false$, $FE_\lambda$ will be written by $E_\lambda$, as shown in Fig.~\ref{fig010}.

\subsection{Future entry index allocation}

\begin{figure}[htbp]
    \centerline{\includegraphics[width=2.7in,trim=0 0 0 0, clip]{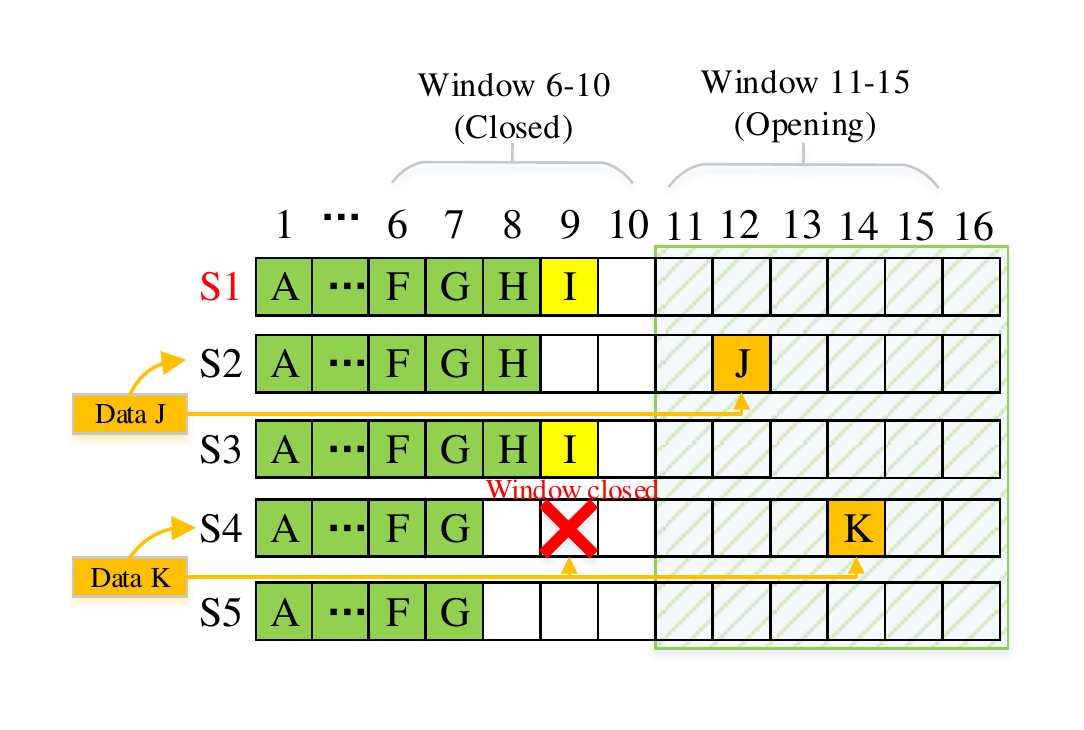}}
    \caption{Example of entry allocation; K will not be written to the position of index 9 because windows 6-10 have been closed when $lastIndex$ of S4 is equal to 6.}
    \label{fig011}
    \end{figure}

We designed a window for LCR to manage the writing process of future entries. A window is a continuous range of index, and it has two states: closed and open. When the window is opened, the node can act as a data-leader to write future entries with an index within its window. When the window is closed, the data-leader will no longer be able to generate future entries with an index in this range. However, for node $S_i$, the state of the window does not reject the entries received from other data-leaders. The condition for the window to close is that the $lastIndex$ of the normal-log (the log that replicated in the sequenced index in SMR) is greater than or equal to the starting index of the window. The window can avoid the conflict described in Section III.B to a certain extent because it can ensure that the allocated index can keep at a distance from the committed index of the leader, as shown in Fig.~\ref{fig011}.
 
In addition, we have added a $generation$ property to the window to ensure that there will be no conflict between future entries when the membership changes. We will introduce it in Section V.

\section{Generation for membership change}

\begin{figure}[htbp]
    \centerline{\includegraphics[width=2.8in,trim=0 18 0 25, clip]{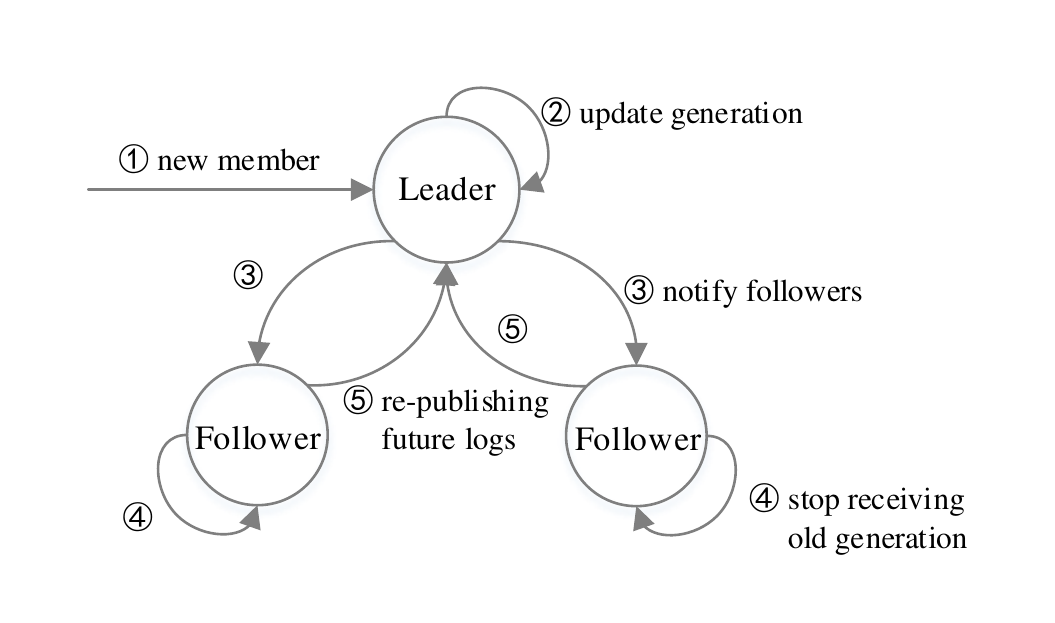}}
    \caption{The state transition diagram of $generation$; the numbers represent the trigger sequence.}
    \label{fig012}
    \end{figure}

In Section IV.A, we mentioned that since the index allocation of the future entries is related to the server id, the dynamic change of the cluster size (membership change) will make the followers meet an index collision. To address the problem, we introduce the $generation$. The $generation$ is an integer that contains information about the number of nodes in the cluster. The $generation$ and $term$ have similar effects. Nodes with higher $generation$ will reject future entries' replication requests from nodes with lower $generation$. At the same time, the node will write its $generation$ in the request and response. When the $generation$ changes, the follower will re-replicate the unapplied future entries that it saves, as shown in Fig.~\ref{fig012}.

In this section, we will explain two key concepts of $generation$: 1) The condition that triggers the $generation$ change. 2) How to re-replicate the future entries for each node to ensure that the entries can keep consistent after the membership changes.

\subsection{Generation changing}

The $generation$ of a node can change under the following three conditions: 1) The node receives the membership changing configuration from the leader. 2) The node learns a newer $generation$ from the replication response. 3) The node learns the new $generation$ from the replication request. When the $generation$ changes, it will immediately stop sending future entries replication requests and re-allocate the index of the future entries that have not been applied to the state machine in such a way that it can be correctly indexed by all nodes. In addition, to facilitate index calculation between nodes, we can use the number of nodes in the cluster as the $generation$. The $generation$’s change rules are flexible. For example, for clusters that frequently change membership, we can take the surplus of $generation$ to reflect the reduction in the number of nodes.

\textit{\textbf{Theorem 2.}} The increase in $generation$ will not cause conflicts between future-logs and normal-logs.

\textit{\textbf{Proof.}} For node $S_i$ with $generation$ $G_i$, future entry $FE_\lambda$ and $\lambda \% G_i= S_i$, when the $generation$ of $S_i$ becomes $G_i'$, $G_i'>G_i$, the new index $\lambda'$ of $FE_\lambda$ is 
\begin{equation}
    \lambda'=\left \lfloor \frac{\lambda}{G_i} \right \rfloor * G_i'+S_i
\end{equation}
Based on Equation (1), $\lambda'$ will not duplicate with other future entries in $G_i'$. In addition, suppose that when the $generation$ increases, the leader's $lastIndex$ of the normal-log is $\omega$. Then, $\lambda > \omega$ because if $\lambda \leq \omega$, $FE_\lambda$ must have been committed in the normal-log according to Section IV.B. Here, we prove that $\lambda' \geq \lambda$ and thus prove $\lambda' > \omega$. From Equation (2), we have the following:
\begin{equation}
    \lambda = k*G_i+S_i
\end{equation}
$\lambda'$ can be presented as
\begin{equation}
    \begin{split}
        \lambda'&=\left \lfloor \frac{\lambda}{G_i} \right \rfloor * G_i'+S_i \\
        &= \left \lfloor \frac{k*G_i+S_i}{G_i} \right \rfloor * G_i'+S_i 
\end{split}
\end{equation}
Since $G_i$ and $G_i'$ are integers, we have $G_i' \geq G_i+1$.
\begin{equation}
    \begin{split}
        \lambda' &\geq \left \lfloor \frac{k*G_i+S_i}{G_i} \right \rfloor * (G_i+1)+S_i \\
        &\geq \frac{k*G_i}{G_i} * (G_i+1)+S_i \\
        &= k * G_i + k + S_i \\
        &\geq  k * G_i + S_i \\
        &= \lambda
    \end{split}
\end{equation}
Therefore, the increase in $generation$ will not cause $FE_\lambda$ to conflict with normal entries because $\lambda$ is larger than any other normal entries. If $FE_\lambda$ can be applied, then $FE_\lambda$ must have been successfully replicated to half of the nodes that contain the leader. This circumstance means that $FE_\lambda$ does not conflict with the normal entries of the leader and thus will not conflict with the normal entries of other nodes, because $\omega$ must be greater than or equal to the $lastIndex$ of the normal-log of all other nodes.

\begin{table*}
    \renewcommand{\arraystretch}{1.1}
    \caption{Configuration settings of the experiments}
    \label{table_3}
    \centering
      \setlength{\tabcolsep}{2.3mm}{ 
      \begin{tabular}{lll}
      \toprule
      Property & Configuration & Comment \\
      \midrule
      election timeout	& 5000 ms & A follower becomes a candidate if it does not receive any message from \\
      &&the leader in election timeout\\
      max await timeout & 1000 ms &The maximum waiting timeout for replicate request.\\
      heartbeat	& 500 ms &A leader sends RPCs at least this often, even if there is no data to send.\\
      max flying requests & 16&Maximum number of concurrent RPC requests.\\
      max entries per request & 5000&Maximum number of entries in each RPC request.\\
      max segment file size	& 100 MB&Maximum number of bytes in each segment.\\
      snapshot period & 60 s&Snapshot interval of each node.\\ 
      \bottomrule
\end{tabular}}
\end{table*}

\subsection{Re-replication in generation changes}

\begin{figure}[htbp]
    \centerline{\includegraphics[width=2.8in,trim=0 10 0 10, clip]{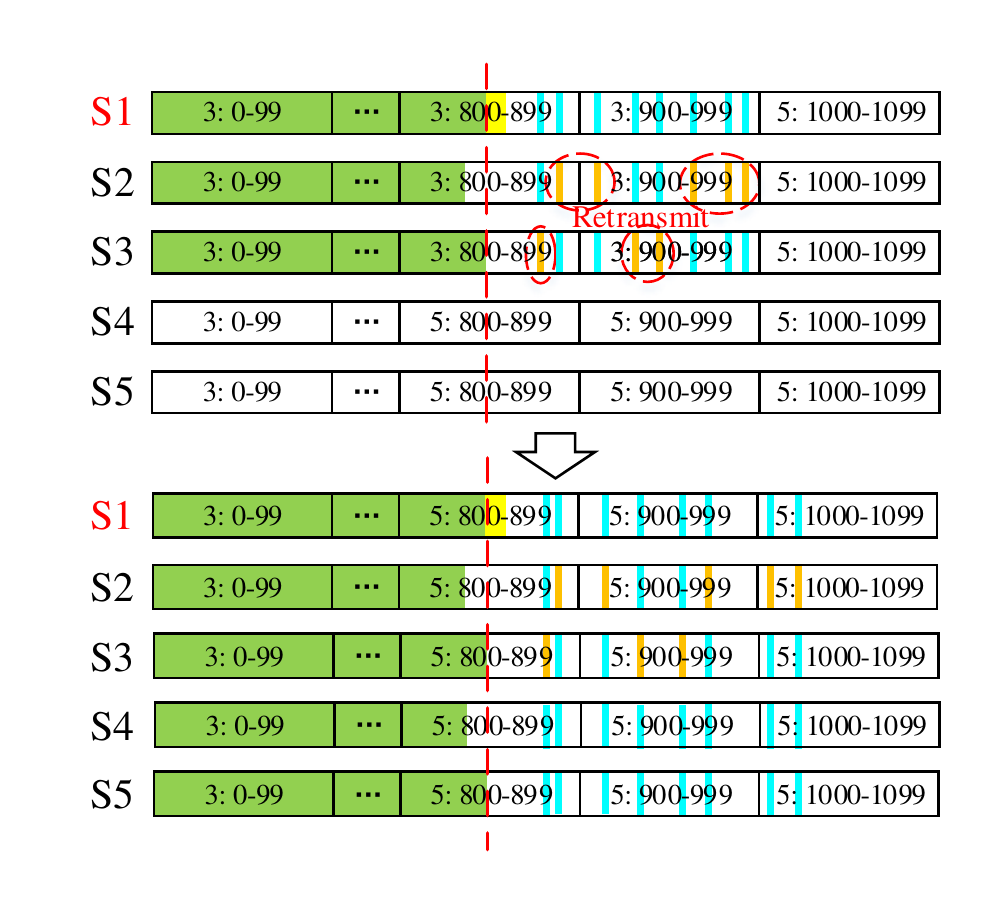}}
    \caption{An example of re-replication in generation changes. The red dashed line is the applied index, and the number format in the box is "$generation$: window range". Greens are applied logs; oranges are future entries that are saved in its data-leader; blues are future entries that are not saved in its data-leader. When generation changes from 3 to 5, each node will re-allocate the index for the future entries and re-replicate to other nodes.}
    \label{fig013}
    \end{figure}

As shown in Section IV, the index allocation method of the future entry is related to the number of nodes. future-log windows with different $generations$ cannot keep the same content in future-logs. Therefore, when the node is notified of the $generation$ changing, it needs to re-allocate and re-replicate the future entries. In detail, when follower $S_i$ obtains the message of $generation$ update, it will immediately record all future entries whose index is larger than the $lastIndex$ of the normal-log, and clear the windows with the old $generation$. At the same time, it creates a new window with updated $generation$, reallocates new indexes for these future entries, and starts to replicate the future entries with the new $generation$. Since the $lastIndex$ of the normal-log on the follower must be less than or equal to the value on the leader, the follower will not conflict with the entries that have been committed in the normal-log, as shown in Fig.~\ref{fig013}.

S1, S2, and S3 store some future entries in two windows: index range 800-899 and $generation = 3$, index range 900-999 and $generation = 3$. When the cluster size changes from 3 to 5, $generation$ is also updated to 5. S2 and S3 need to recreate the window with $generation = 5$. In the window where $generation=3$, the nodes will reallocate the index of future entries in it to the new window according to Equation (1) . 

\textit{\textbf{Theorem 3.}} After the generation is updated, the re-replication process will not cause inconsistencies in the future-log.

\textit{\textbf{Proof. }} In node $S_i$, for an unapplied future entry $FE_\lambda$ with generation $G_i$, $\lambda\%G_i=S_i$, and $lastIndex_i$ of the normal-log is $\omega$, $\omega<\lambda$. When $generation$ is updated to $G_i'$, the new index of $FE_\lambda$ is $\lambda'=\left \lfloor \frac{\lambda}{G_i} \right \rfloor * G_i'+S_i$. In this case, node $S_i$ will replicate $FE_\lambda'$ to all other nodes with generation $G_i'$. If for node $S_j$, its $lastIndex_j$ of the normal-log is greater than or equal to $\lambda'$, it will ignore $FE_\lambda'$, which also indicates that the leader’s $lastIndex$ of the normal-log is greater than $\lambda'$. Therefore, $FE_\lambda'$ cannot be applied since it will not be committed by quorum. Node $S_i$ needs to re-allocate the index of $FE_\lambda'$ according to the re-replication strategy in Section IV.B. If for node $S_j$, its $lastIndex$ of normal-log is less than $\lambda'$, it will accept $FE_\lambda'$ on the premise of confirming that its $G_j = G_i'$. According to \textit{Theorem 1}, the acceptance of $FE_\lambda$ will not cause conflict.

\section{Experiments and analysis}

Due to the complexity of Paxos implementation, Raft has become the main consensus protocol adopted by the private chain due to its advantages in implementation difficulty and efficiency. Therefore, we implemented LCR based on Raft and evaluated it under different network latencies and node numbers to test its scalability and robustness. First, we experiment on the correctness of LCR-Raft to verify its protocol's functional support for native Raft. Second, we evaluate the performance of LCR-Raft to verify its improvement in specific business scenarios. Finally, we compare the optimization effect of LCR-Raft on the leader node in terms of the CPU and network traffic consumption with native Raft and multi-Raft.

\subsection{Environments and configurations}

In order to verify the performance improvement effect of LCR-Raft, we implemented it in the open-source Raft consistency model\cite{raftjava}. Among them, the RPC protocol is implemented using brpc\cite{brpc}, and RocksDB\cite{rocksdb} 5.1.4 is used to store state machine data. All of the instances are deployed on 3, 5, 7, and 9 servers (according to the number of nodes in the different experimental configurations) with a Xeon E5-2620 v4 CPU, 64 GB of RAM, 4 TB, and 7200 RPM SATA disk. The system runs CentOS Linux release 7.5.1804, 64 bit, with Linux kernel Version 3.10.0 The switch uses H3C Mini S8G-U, with 16 Gbps switching capacity and 12 Mpps forwarding performance.

Our client runs on 2 servers with 6 core 12 thread AMD 3600 CPU and 32 GB memory, and the default number of client jobs is 40. The non-transactional data are the real-time sensor data generated from our blockchain water quality monitoring project, while the transactional data are value transferring operations. The default transactional data and non-transactional data size are 80 bytes. The base configuration of Raft, LCR, and multi-Raft is detailed in TABLE~\ref{table_3}. 

It is worthwhile to note that to compare LCR-Raft and other consistency models fairly, we use pipelines to implement SMR, i.e., each log contains only one transaction, and the consensus node immediately invokes remote replication when it generates future entries. The reason is that our work focuses on the optimization of SMR. Although the use of batch messages could improve TPS tens to hundreds of times, its effect is related to the number of clients, the request frequency, the batch size, and the dynamically adjusted request queue, which makes it unable to truly reflect the contributions of our work.

\subsection{Response time}

LCR decouples the replication process of future-log and normal-log. Therefore, for transactional data, the client can still use a normal-log to submit the transactional data. For non-transactional data, using a future-log can significantly reduce the leader's overhead and the response time of normal entry requests (because normal entries will be processed immediately by the leader). Therefore, in this section, we analyze and test the latency of applying transactional and non-transactional data, and compare it with Raft to verify its optimization effect.

For non-transactional data $D_{nt}$, the latency $\Delta_{nt}$ is 
\begin{equation}
    \Delta_{nt} = 2\Delta_c + 2\Delta_n + \eta_f + \eta_w + \iota 
\end{equation}
where the $\Delta_c$ is the network latency between the cluster and client, $\Delta_n$ is the network latency between consensus nodes in the cluster, $\eta_l$ and $\eta_f$ are the time periods from entry packaging to replication request sending on the leader and followers, and $\eta_w$ is the time period during which the data-leader wait for a signal entry to apply the data. Here, $\iota$ is composed of the disk operation latency, such as log writing and state machine applying. 

For transactional data $D_{t}$, the latency $\Delta_{t}$ is 
\begin{equation}
    \Delta_{t} = 2\Delta_c + 4\Delta_n + \eta_l  + \iota
\end{equation}
In $\Delta_{nt}$ and $\Delta_{t}$, $2\Delta_c$ is generated from the network latency between client the and nodes. As shown in Figure. \ref{fig006}, it needs $2\Delta_n$ for $D_{nt}$ and $4\Delta_n$ for $D_{t}$ to respond to the clients.

For $\Delta_n$, since the clients do not know which node is the leader, both transactional data and non-transactional data can be sent to any node. When transactional data are sent to a follower, the follower redirects the request to the leader. In this case, an additional RTT is generated from the redirection. Since this case occurs at a $\frac{N-1}{N}$ probability, the expected latency is $\frac{4N-2}{N}\Delta_n \approx 4\Delta_n$. The non-transactional data need $2\Delta_n$ to respond to clients since they will not be redirected.
\begin{figure*}[htbp]
    \centerline{\includegraphics[width=5.9in,trim=0 0 130 320, clip]{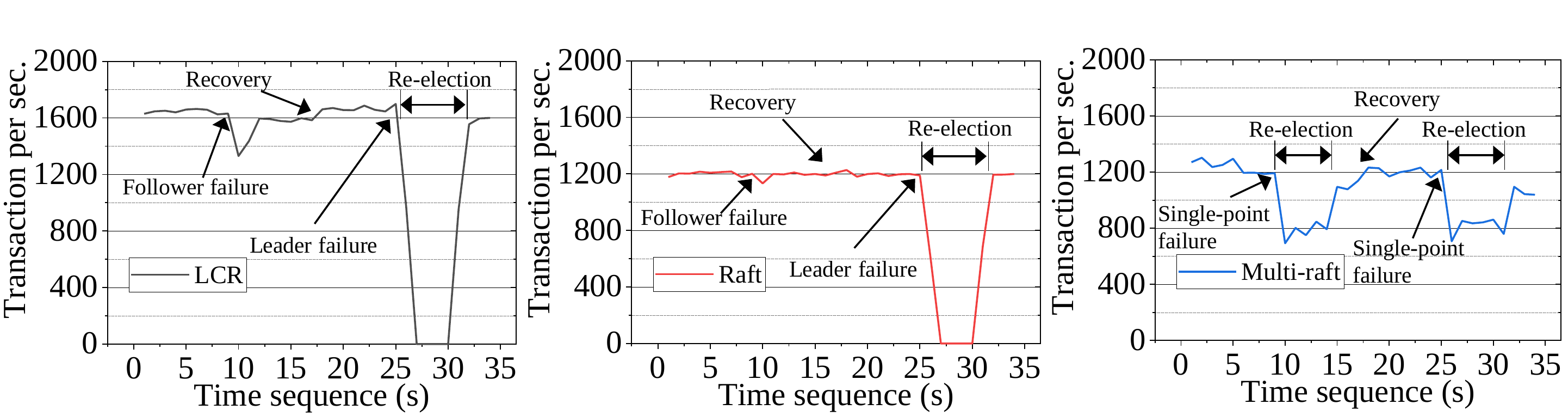}}
    \caption{TPS performance of LCR, Raft and multi-Raft when facing node failure. The left subfigure is the LCR, the middle subfigure is Raft, and the right side is multi-Raft.}
    \label{fig015}
\end{figure*}
In addition to the $\Delta_n$, the differences in latency for $\Delta_{nt}$ and $\Delta_{t}$ are $\eta_w$, $\eta_l $, and $\eta_f$ since $\Delta_c$ and $\iota$ are fixed latencies. From the experiments in Figure. \ref{fig004}, it can be found that the additional latency is generated from $\eta_l$ (tens-ms) because all other conditions are the same besides the request frequency from the clients. The limited worker threads and high network latency make $\eta_l$ higher as the client increases. Similar to $\eta_l$, $\eta_f$ increases because the high request frequency of non-transactional data makes followers run under high loads. Here, $\eta_w$ could be higher than 1 RTT since the leader might not send signal entries to the data-leader immediately when it receives future entries. Besides, the overhead on the leader could amplify $\eta_w$. However, $\eta_l$ will decrease significantly because the proportion of normal entries decreases. It allows the leader to reduce the number and frequency of normal entry replication operations. Therefore, the response time of the transactional data in LCR is lower than in Raft.

\begin{figure}[htbp]
    \centerline{\includegraphics[width=3.2in,trim=0 10 485 0, clip]{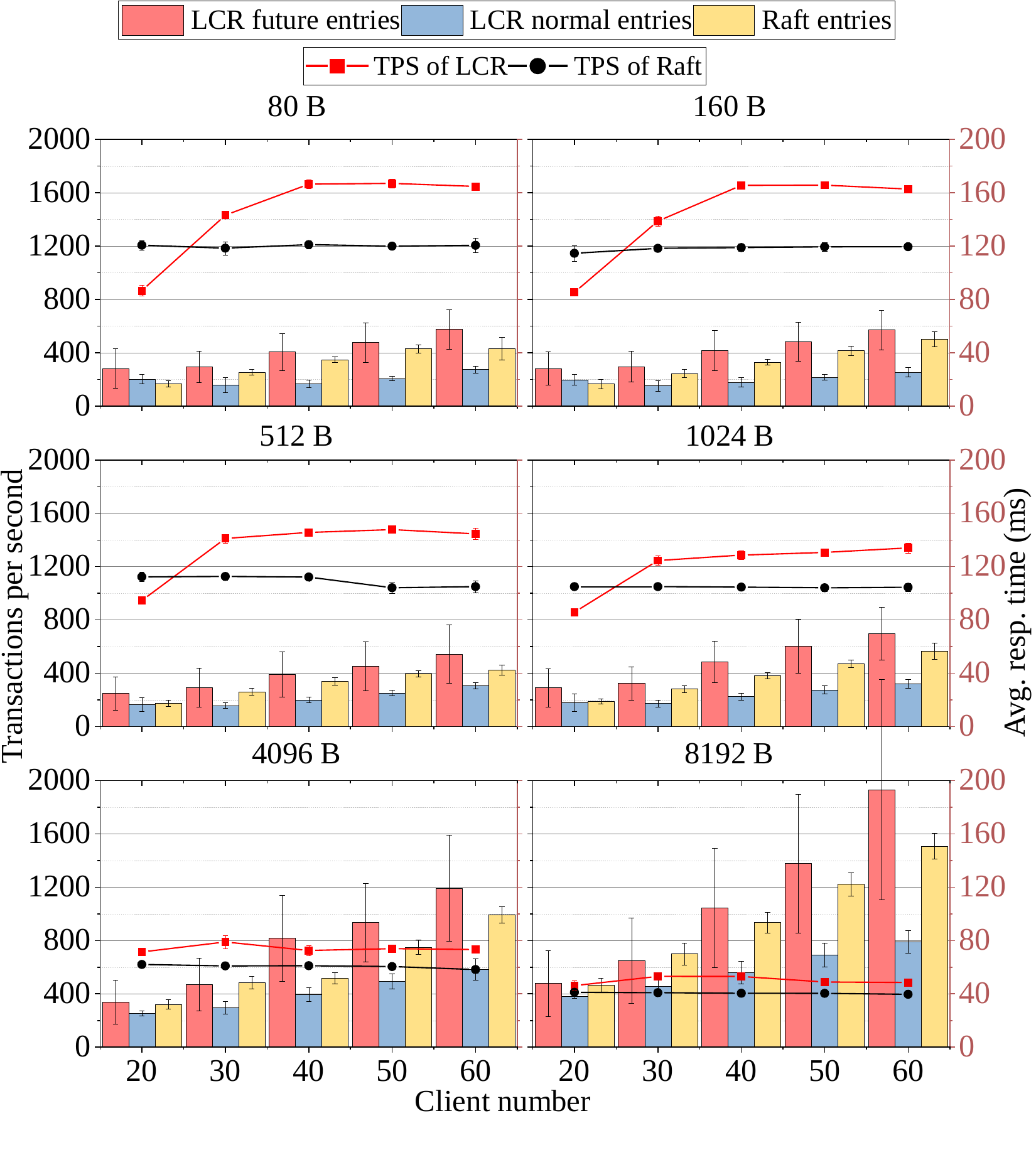}}
    \caption{Comparison between LCR and Raft in TPS and the response time with different request sizes and client numbers.}
    \label{fig014}
    \end{figure}

We tested the effect of LCR and Raft on TPS and the response time in a 5-node, 5 ms latency environment, which is the common network latency between distributed blockchain systems. The proportion of non-transactional (LCR future entries) data was 25\%. The result is shown in Fig.~\ref{fig014}.

In terms of the TPS, LCR outperforms approximately 30\% when the client count is larger than 20. For the response time, as the number of clients and the entry size increase, resource starvation becomes more serious. This circumstance makes the effect of the LCR on reducing the concentrated load more obvious. LCR can provide a 40\%-60\% response time decrease for transactional data (LCR normal entries), which matches the observation in Fig.~\ref{fig003}.

\subsection{LCR re-election}

To show the availability of LCR, we test the node failures in the cluster. In this experiment, LCR, Raft, and multi-Raft run in a 5-node, 5 ms network latency environment. We randomly stop a non-leader node (an arbitrary node for multi-Raft) at the 10th second and reconnect it to the cluster at the 16th second. In the 25th second, we stop the leader node (an arbitrary node for multi-Raft) and no longer recover it. The result is shown in Fig.~\ref{fig015}.

\textit{LCR}: When the data leader is offline, the future entries that it generates might have not been replicated to all nodes. In this case, the followers that did not receive these future entries will ask the leader to re-replicate these entries, since the leader uses signals to replicate them (as described in Section IV.B). This circumstance will destroy the performance improvement brought by batching. However the performance can be quickly restored when these entries are handled. Nonetheless, the TPS will decrease slightly before the node recovers since the downtime of the followers reduces the frequency of future-entry generation, which reduces the optimization effect of LCR. During the 10th second to the 16th second, other followers can still act as data-leaders to replicate future entries. Therefore, the performance is not adversely affected (from approximately 1700 TPS to 1600 TPS). When the leader node fails (at the 25th second), it will trigger re-election, which makes the service unavailable during this period. When the new leader is elected (at the 31st second), the TPS performance can recover.

\textit{Multi-Raft}: In multi-Raft, every node is the leader of shards. For a shard $C_i$ whose leader is $S_i$, failure on $S_i$ will make $C_i$ unavailable because of re-election. Although the TPS does not drop to 0 (from approximately 1200 TPS to 800 TPS), the remaining TPS is contributed by the active shards. Therefore, the performance of multi-Raft is more sensitive to node failures, since when any node fails, performance fluctuations will occur. However, in LCR and Raft, a non-leader node failure will not lead to serious performance loss. In addition, the availability problem still exists in multi-Raft. Since the probability of each node being offline is the same, the total availability performance is the same as that of Raft.

\textit{Raft}: Raft suffers the least performance loss when facing a failure of the follower node. However, due to the high load on the leader, its performance under a 5 ms network latency is only approximately 60\% (approximately 1200 TPS to 1700 TPS) of the LCR.

\subsection{LCR performance experiments}

As an important performance indicator of the SMR protocol, we tested the TPS of LCR. In addition, we conducted experiments on the CPU load and network traffic to verify its effectiveness in reducing the load of the leader.

\subsubsection{TPS}

\begin{figure}[htbp]
\centerline{\includegraphics[width=3.7in,trim=0 0 195 30, clip]{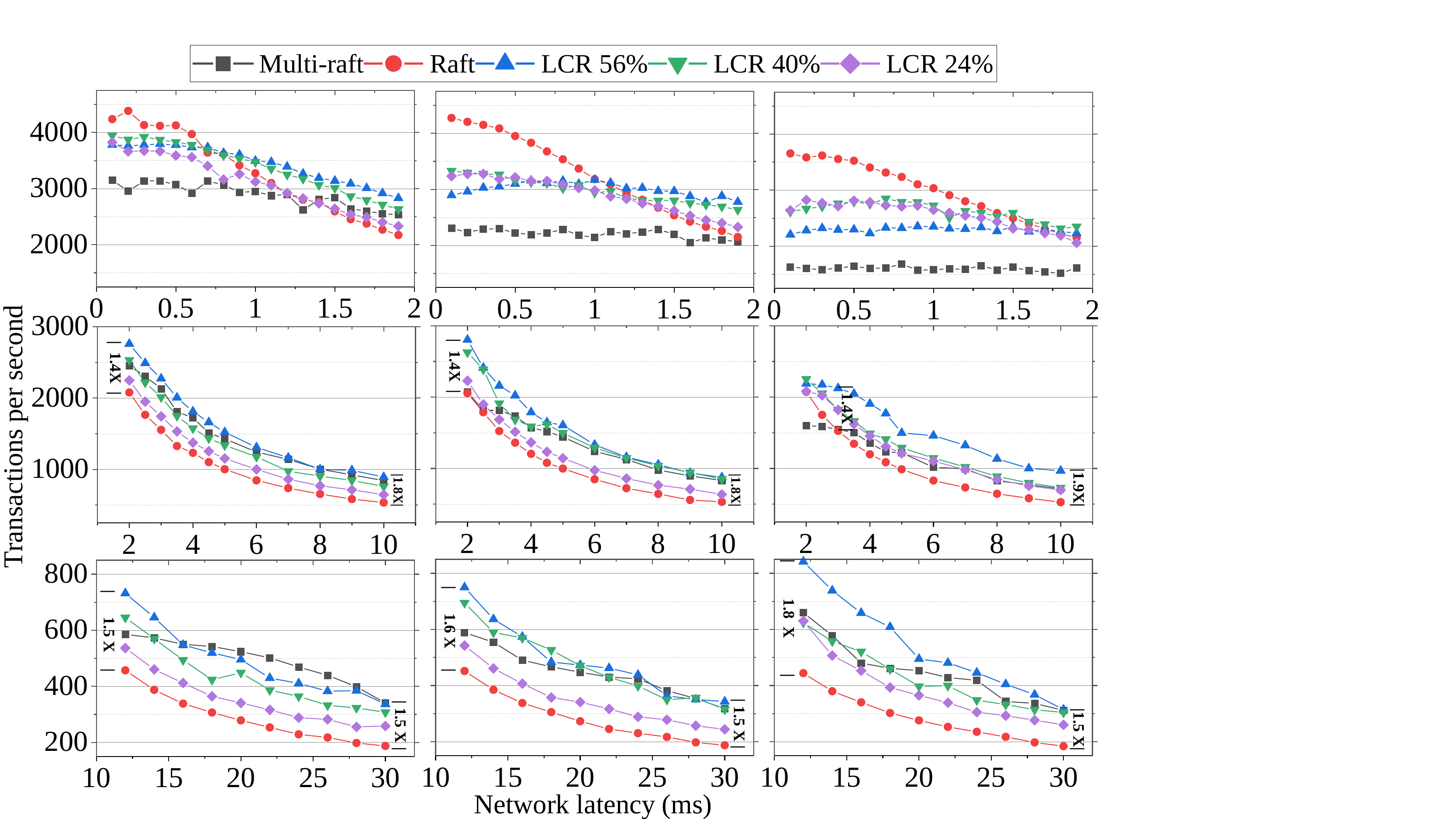}}
\caption{TPS comparison of LCR, Raft and multi-Raft in 5, 7 and 9 nodes and different network latency environments.}
\label{fig016}
\end{figure}

We tested the performance of LCR in terms of the TPS under different cluster sizes and different network latency environments. We designed three clusters with node numbers of 5, 7, and 9. In terms of the network latency setting, we use the $tc$ command of Linux to set the simulated delay on the network card and design 41 test points, which are 0, 0.1, 0.2, 0.3, 0.4, 0.5, 0.6, 0.7, 0.8, 0.9, 1, 1.1, 1.2, 1.3, 1.4, 1.5, 1.6, 1.7, 1.8, 1.9, 2, 2.5, 3, 3.5, 4, 4.5, 5, 6, 7, 8, 9, 10, 12, 14, 16, 18, 20, 22, 24, 26, 28, and 30. The probability of the network latency fluctuation is 30\%, 0.1 ms. We compared LCR-Raft with Raft and multi-Raft. Among them, we optimized multi-Raft using RPC reuse to improve the communication efficiency. On the client side, we created 40 clients to compete to call the RPC protocol to make the cluster work under the maximum load. For LCR, we designed 3 workloads with different non-transactional data (future entries) ratios, LCR 24\%, LCR 40\%, and LCR 56\% to show the effect of the proportional relationship between the two on the performance of the mechanism. The experiment results are shown in Fig.~\ref{fig016}.

To more clearly show the performance of LCR under different network latencies, we divide the performance comparison chart into three intervals (low latency, medium latency, and high latency) at 2 ms and 10 ms. When the network latency is of medium latency and high latency, the LCR has a significant improvement compared with the Raft of approximately 1.4X-1.9X. Under the workload with a higher proportion of future entries, the LCR also has a significant improvement compared to multi-Raft. In the low latency interval, Raft has better TPS performance when the latency is less than 1 ms. Therefore, the effect produced by our model is more suitable for geo-distributed systems with ms-network latency.

\subsubsection{CPU loads}

\begin{figure}[htbp]
\centerline{\includegraphics[width=2.8in,trim=0 0 750 355, clip]{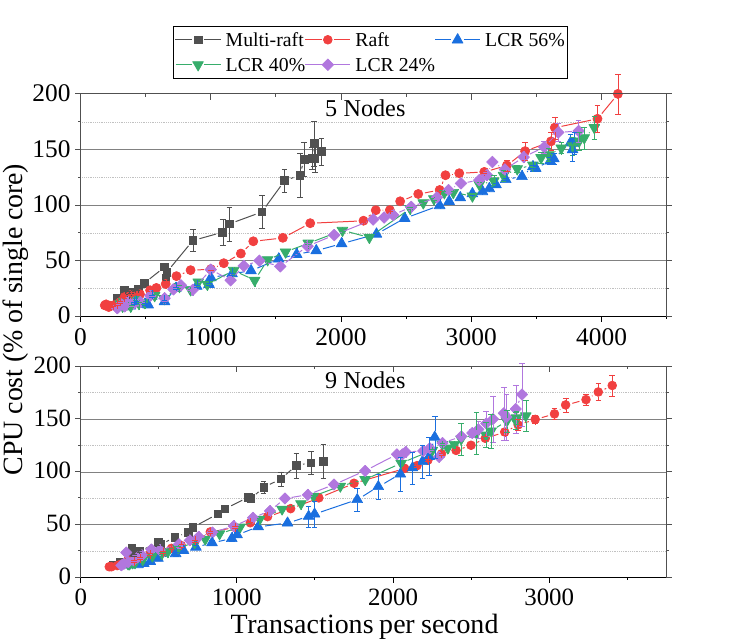}}
\caption{CPU cost in 5-node and 9-node environments (as a percentage of the single-core max load).}
\label{fig017}
\end{figure}

Since the replication process of the future-log is dominated by followers, it will increase the CPU load on the followers in LCR. Although the leader does not need to control the replication process of future-logs, it still needs to handle signal replacement and retransmission. Therefore, LCR can also generate CPU overhead to the leader. As shown in 6.1, under the same network latency, different consistency protocols have obvious gaps in the TPS. Considering that it is unfair to directly compare the CPU load when the network latency is the same, we use TPS as an x-axis to observe the CPU load of the consistency protocols when they have the same TPS. The experiment results are shown in Fig.~\ref{fig017}.

The CPU load of the LCR leader node is similar to that of Raft. When the number of nodes is 5, the CPU load of LCR is slightly higher than that of Raft. When the number of nodes is 9, the CPU load is reduced by 5\%-10\%.

\subsubsection{Network traffic}

\begin{figure}[htbp]
\centerline{\includegraphics[width=2.75in,trim=0 0 775 300, clip]{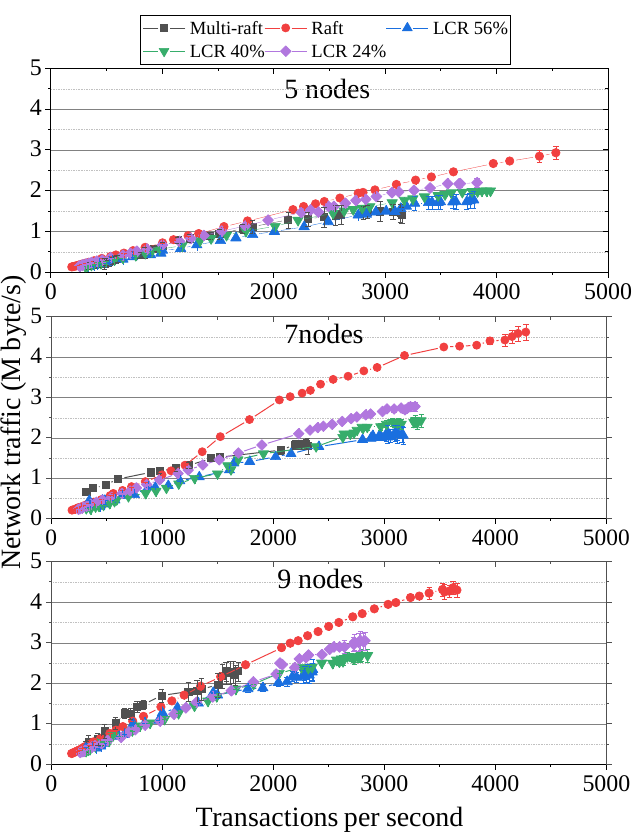}}
\caption{Comparison figure for the network traffic consumption in a 5-node, 7-node, and 9-node environment.}
\label{fig018}
\end{figure}

As with the CPU load, we also use TPS as a variable to compare their network traffic. Since LCR transfers most of the log replication work to followers, when the TPS is the same, the leader node will generate smaller network traffic. Similarly, the degree of network traffic savings is also related to the non-transactional data ratio. Therefore, we computed independent statistics on the LCR traffic under different non-transactional data ratios. The experiment results are shown in Fig.~\ref{fig018}.

LCR can save 20\%-30\% of the network traffic of the leader node. We can conclude that LCR achieves the same TPS as Raft with a lower throughput rate, which can effectively provide saving in the costs of network construction.

\subsection{Overhead analysis}

LCR can generate additional resource costs according to the following three points. 1) Network traffic: This cost arises because the follower nodes undertake part of the log replication work. As a result, the network traffic is increased in the followers. 2) Memory consumption: the leader needs memory to cache the relational mapping to ensure the writing efficiency of the future-log. 3) Computational resource consumption: future-log replication requires index calculation and verification. We calculated statistics on these costs to measure whether the extra cost generated by LCR is acceptable.

\begin{figure}[htbp]
\centerline{\includegraphics[width=2.8in,trim=0 0 775 300, clip]{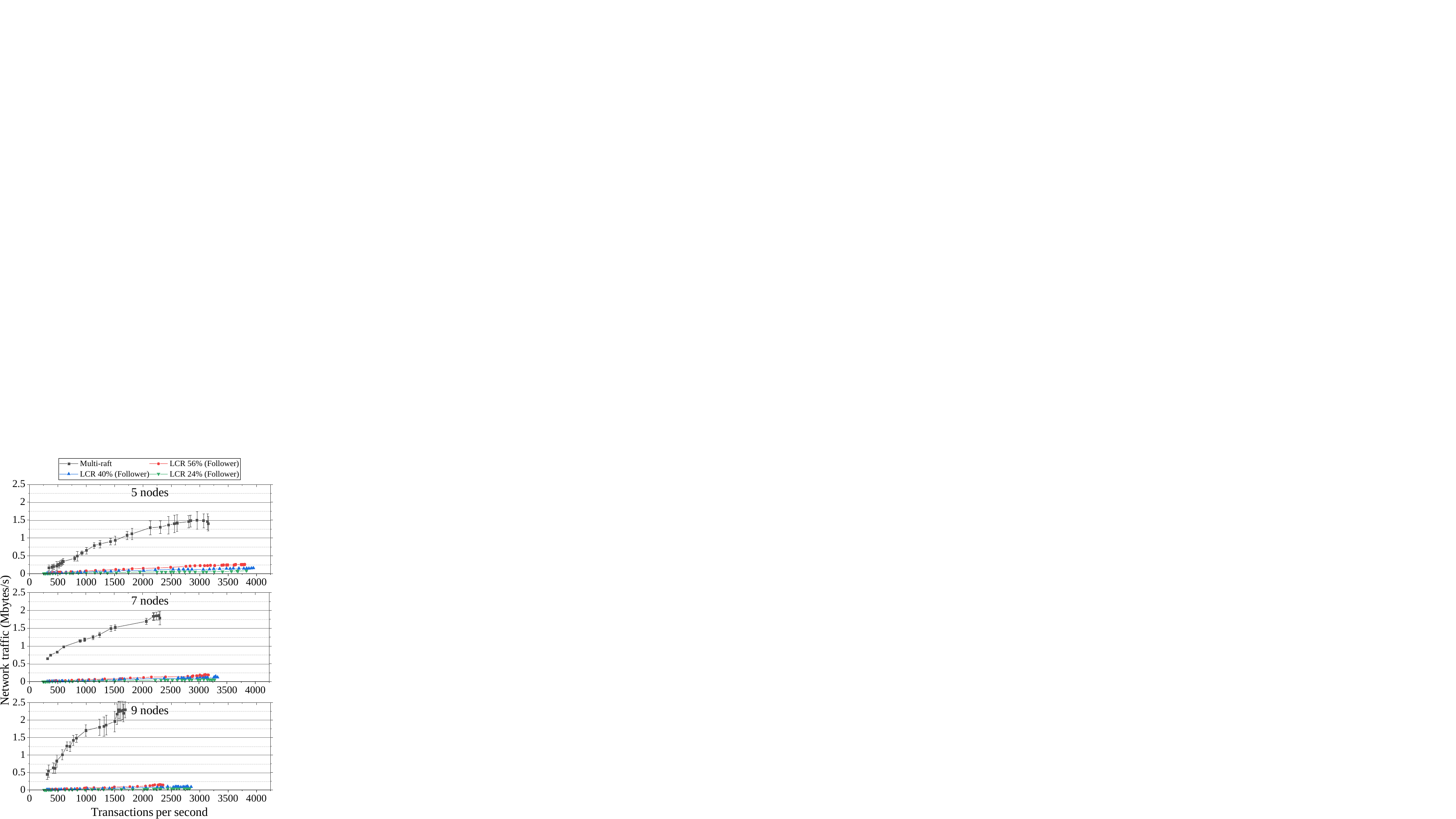}}
\caption{Additional network traffic from followers.}
\label{fig019}
\end{figure}

First, we measure the network traffic. Since additional network traffic is generated on the follower nodes, we will not compare the leader's network traffic. Here, we count the extra network traffic with different proportions of future entries and compare it with the traffic consumption of multi-Raft. The traffic consumption of multi-Raft is the average of all of the nodes. Similar to Section VI.C, we still compare the traffic consumption under the same TPS. The experiment results are shown in Fig.~\ref{fig019}. The additional network traffic generated by LCR is much less than that generated by multi-Raft.

The consumption of the network traffic generated by LCR is much less than that of multi-Raft traffic. At the same time, we found that at the same TPS, the network consumption of the followers will decrease as the cluster size increases. This finding is in line with our expectations because each follower needs to replicate fewer future entries.

\begin{figure}[htbp]
\centerline{\includegraphics[width=2.55in,trim=0 0 800 460, clip]{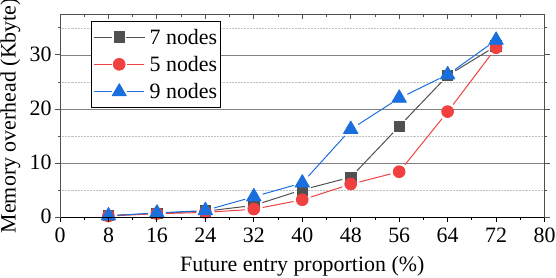}}
\caption{Additional memory overhead from the odes in LCR. The experiments run in the 5-, 7-, and 9-node clusters, the average network latency is 5 ms, and the average size of key-value pairs is 80 bytes.}
\label{fig020}
\end{figure}

In the consistency protocol, the entries must be written to the segmented log and memory at the same time before it is applied to the state machine to ensure that the received log will not be lost. Therefore, future entries need to consume additional memory. We tested the LCR on the extra main memory consumption. The experiments were run in an environment in which the number of consensus nodes is 5, 7, and 9, the average network latency is 5 ms, and the average size of key-value pairs is 80 bytes. The experiment results are shown in Fig.~\ref{fig020}.

The experiments show that the consumption of LCR on the main memory is acceptable (KB level) and that the proportion of future entries is positively related to the additional consumption of the main memory. When the number of future entries is too large, the growth rate of its index will far exceed that of normal entries. However, applying the future entries to the state machine must wait for the normal entries whose index is greater than it to be applied. In extreme cases, such as when the proportion of future entries reaches 99\%, too many future entries will make the growth rate of the normal entries index unable to catch up with the future entries index. This circumstance will cause future entries to wait a very long time before it can be applied. To address this problem, we designed a stepping mechanism to allow the leader to confirm future-entries in advance.

\begin{figure}[htbp]
    \centerline{\includegraphics[width=3in,trim=0 0 740 435, clip]{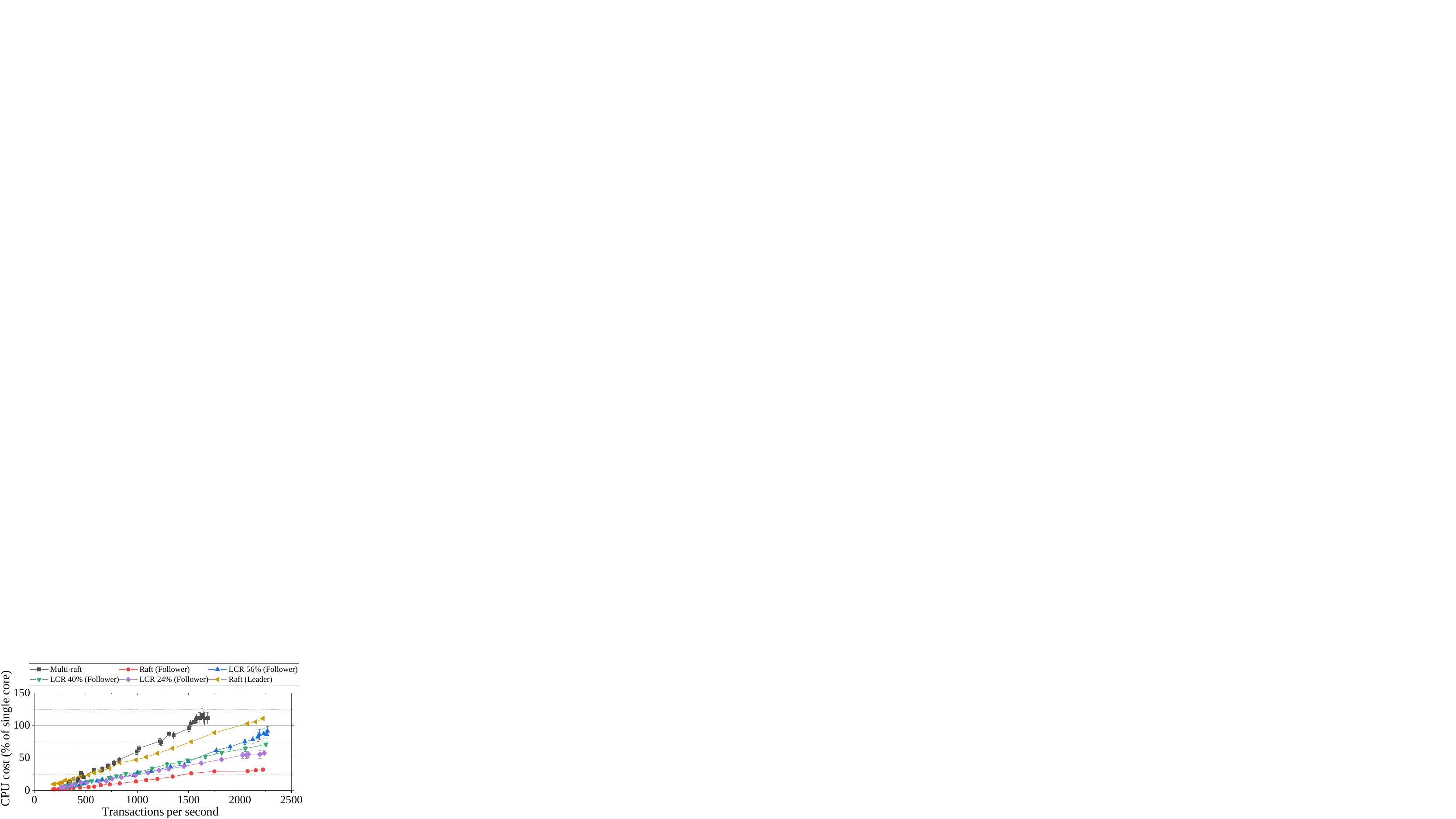}}
    \caption{CPU consumption of nodes with multi-Raft, Raft leader, Raft follower and LCR.}
    \label{fig021}
    \end{figure}

For computational resources, LCR only needs extremely low complexity such as list and hash map lookup. Although we calculated statistics on leader resource consumption in Section VI.C, our model did not cause a large overhead on the leader node. However, our model will generate an additional CPU load on the follower nodes. We calculate statistics on the CPU load of followers on Raft, multi-Raft, and LCR under different TPSs in a 9-node cluster in Fig.~\ref{fig021}.

The results show that the CPU load of LCR will not exceed the leader in Raft. This finding is acceptable since each node should have similar processing capabilities to be elected as the leader. Therefore, the CPU load of LCR will not cause performance degradation.

\section{Discussion}

\subsection{Confirmation vs. multi leader}

LCR is an SMR model based on leader confirmation, but it performs better than semi-synchronous in security. LCR did not decentralize the power of the leader. There is a large number of existing methods that use multi-leader to reduce the load on the leader, such as multi-Raft\cite{huang2020tidb} and Hash Raft\cite{yang2019blockchain}, etc. However, regardless of the implementation, it will face the following two problems: 1) When any node becomes unavailable, the cluster must invoke re-election. 2) Each pair of nodes must synchronize their online status through heartbeats. The leader confirmation method solves these two problems well. First, when the follower node is unavailable, it cannot copy and receive the log from followers and the leader. This circumstance does not require the re-election process. At the same time, compared with multi-Raft, it does not require heartbeats to maintain the online status of each node. Because it does not need the followers to know whether other nodes are online, the follower only needs to know whether the future-logs are accepted by half nodes that contain the leader. Therefore, it does not need to reuse the connection to reduce the network load between nodes, which greatly reduces the complexity of the protocol. In addition, LCR does not guarantee that every future-log has been written into the state machine when it returns a success message to the client. However it will ensure that data will be written to the state machine at some point in the future, which also makes it applicable to more complex network environments.

\subsection{More than Raft}

LCR is an SMR model based on leader confirmation, but it performs better than semi-synchronous in terms of consistency. LCR did not decentralize the power of the leader. There is a large number of existing methods that use multi-leader strategies to reduce the concentrated load\cite{huang2020tidb,yang2019blockchain}. However, regardless of the implementation, it will face the following two problems: 1) When any node becomes unavailable, the cluster must invoke re-election. 2) Each pair of nodes needs to synchronize their online status through heartbeats. LCR solves these two problems well. For the follower node failure cases, the node cannot copy or receive the log from followers and the leader. It does not require re-election. At the same time, compared with multi-Raft, LCR does not require heartbeats to maintain the online status of each node. Because followers do not need to know whether other nodes are online, the only thing they care about is whether the future-entries are confirmed by the leader, which greatly reduces the complexity of the protocol.

\subsection{Cost}

The improvement brought by the LCR does not cause additional costs. As described in Section VI.D, the overhead does not cause consumption that exceeds the original system capacity. Because the log replication process of the SMR protocol is extremely concise, the network, CPU, and other overhead are very sensitive to consensus protocols. As we described in Section II, the increase in the latency will make the TPS drop significantly. To reduce the network latency between geo-distributed nodes, it might be necessary to establish a dedicated network, which is very difficult in terms of feasibility. Although the cost of upgrading the CPU and router is lower, it can also improve the optimization effect brought by the LCR. Therefore, the LCR can trade at a lower cost for the improvement of transactional data consistency efficiency and TPS. In addition, as described in Section VI.C, the LCR can reduce substantially the network traffic. For cluster deployment in WAN, network bandwidth or traffic must be purchased from the Internet Service Provider (ISP), and the cost is positively related to the bandwidth or traffic. Therefore, the LCR can effectively save approximately 20\%-30\% of the network construction costs.

\section{Conclusions}

In this paper, we propose LCR for geo-distributed consensus and non-transactional data. A novel leader confirmation mechanism is proposed, which can reduce the leader's concentrated load. LCR implements a future-log on the SMR protocol and proves its availability. It makes LCR transparent to the election mechanism, which can support various consensus protocols, and combine with other optimization schemes. We implemented LCR based on Raft and tested it in clusters with 3, 5, 7, and 9 nodes. Experimental results show that LCR improves the TPS at 1.4X-1.9X that of Raft, reduces the transactional data response time by 40\%-60\%, and the network traffic by 20\%-30\% when the network latency is greater than 1.5 ms compared with Raft. In addition, the overhead on the followers is much lower than that of multi-Raft, and the CPU consumption on the followers is lower than that of the leader node in Raft.

\section*{Acknowledgment}
This work is supported by the Fundamental Research Funds for the Central Universities (grant no. HIT.NSRIF.201714), Weihai Science and Technology Development Program (2016DXGJMS15), and the Key Research and Development Program in Shandong Province (2017GGX90103).

\bibliographystyle{IEEEtran} 
\bibliography{IEEEabrv, bib}  

\end{document}